\documentclass[10pt,a4paper,3p]{elsarticle}

\usepackage{mathpazo}
\usepackage[T1]{fontenc}
\usepackage[latin9]{inputenc}
\usepackage{color}
\usepackage{amsmath}
\usepackage{amssymb}
\usepackage{graphicx}
\usepackage[unicode=true,
 bookmarks=true,bookmarksnumbered=false,bookmarksopen=false,
 breaklinks=false,pdfborder={0 0 0},pdfborderstyle={},backref=false,colorlinks=true]
 {hyperref}
\hypersetup{
 colorlinks,citecolor=blue,filecolor=black,linkcolor=red,urlcolor=blue}
\usepackage{breakurl}

\makeatletter

\usepackage{color}
\biboptions{sort&compress}
\journal{arXiv}

\makeatother

\begin{document}

\title{Modulation of localized solutions in quadratic-cubic nonlinear Schr\"odinger
equation with inhomogeneous coefficients}

\author{Wesley B. Cardoso$^{a,}$\footnote{Corresponding author. Tel.: +55 6235211014.\\
E-mail address: wesleybcardoso@gmail.com (W. B. Cardoso).}, Hugo L. C. Couto$^{a}$, Ardiley T. Avelar$^{a}$, Dionisio Bazeia$^{b}$}

\address{$^{a}$Instituto de F\'{i}sica, Universidade Federal de Goi\'as, 74.690-900,
Goi\^ania, Goi\'as, Brazil}

\address{$^{b}$Departamento de F\'{i}sica, Universidade Federal da Para\'{i}ba,
58051-970 Jo\~ao Pessoa, Para\'{i}ba, Brazil}
\begin{abstract}
We study the presence of exact localized solutions in a quadratic-cubic
nonlinear Schr\"odinger equation with inhomogeneous nonlinearities.
Using a specific \emph{ansatz}, we transform the nonautonomous nonlinear
equation into an autonomous one, which engenders composed states corresponding
to solutions localized in space, with an oscillating behavior in time.
Direct numerical simulations are employed to verify the stability
of the modulated solutions against small random perturbations. 
\end{abstract}
\maketitle

\section{Introduction}

Localized solutions in nonlinear media such as the shape preserving
solitons, and breathers, which are characterized by internal oscillations,
have fundamental applications for energy transport in optical fibers
and waveguides \cite{Agrawal_13,Flach_PR98}, polaronic materials
\cite{Andersen_PRB93}, biological molecules \cite{Peyrard_PD93},
etc. These solutions can propagate without losing their shape due
to equilibrium between diffraction (in spatial domain) or dispersion
(in temporal domain) and nonlinearity \cite{Scott_03,Drazin_89,Remoissenet_13,Eilenberger_12}.
They also appear describing localized excitations in dilute Bose-Einstein
condensates in specific conditions of balance between the particle
dispersion and the nonlinear effect due to the two-body interaction
\cite{Pethick_02,Pitaevskii_03} and open possibilities for future
applications in coherent atom optics, atom interferometry, and atom
transport.

In many cases, the equations that describe the systems have the form of a
nonlinear Schr\"odinger (NLS) equation \cite{Malomed06-2}. Indeed,
the NLS equation appears as a universal equation governing the evolution
of slowly varying packets of quasi-monochromatic waves in weakly nonlinear
media featuring dispersion. One can exemplify this with: light propagation in nonlinear
optical fibers and planar waveguides; small-amplitude gravity waves
on the surface of deep inviscid water and Langmuir waves in hot plasmas;
interaction of high-frequency molecular vibrations and low-frequency
longitudinal deformations in a model of long biological molecules;
mean-field description of Bose-Einstein condensates (in this
case it becomes the Gross-Pitaevskii equation); models
that describe nonlinear dissipative media (in this case it becomes the Ginzburg-Landau
equation); and so on. Regarding the form of the nonlinearities in
the NLS equation one can find: quadratic \cite{Buryak_PR02}, quadratic-cubic
\cite{Hayata_JOSAB94,Fujioka_CHAOS11}, cubic (or Kerr-type) \cite{Agrawal_13,Dalfovo_RMP99,Aranson_RMP02},
cubic-quintic \cite{Akhmediev_PRE96,Mihalache_PRE00,Avelar_PRE09,Cardoso_PRE11},
quintic \cite{Alfimov_PRA07,Reyna_PRA14}, nonpolynomial \cite{Salasnich_PRA02,Salasnich_PRA02-2,Mateo_PRA08,Adhikari_NJP09,Cardoso_PRE13,Couto_JPB15},
logarithm \cite{Biswas_CNSNS10,Calaca_CNSNS14}, saturable \cite{Soto-Crespo_PRA92,Stepic_PRE04,Melvin_PRL06},
and other nonlinearities.

The inclusion of variable coefficients in the NLS equation, which
are due to inhomogeneities, produces a gigantic backdrop of possibilities
of modulation of the localized solutions, although it breaks the integrability
of the standard cubic NLS equation \cite{Yang_10}. Physically, the
variable coefficients can be achieved by changing the system structure,
as for example, in the magnetohydrodynamics the inhomogeneity of the
real plasma environment can be achieved by fluctuations of the density,
temperature, and magnetic fields \cite{Wang_PP15}; the inhomogeneities
in nonlinear fibers or crystals are due to variation in the geometry
and/or variation in the material parameters in the fabrication process
of such systems \cite{Agrawal_13}; in Bose-Einstein condensates (BECs)
the variations in the potential and nonlinearities can be controlled
by the application of external fields, which also induce modulation
pattern of the local nonlinearity through the Feshbach-resonance mechanism,
i.e., field-induced changes of the scattering length characterizing
binary collisions between atoms, which contributes to modify the nonlinearity
in the BEC \cite{Theis_PRL04}.

The construction of analytical solitonic solutions is a hard task.
However, by applying a similarity transformation technique, which
transforms the nonautonomous NLS equation into an autonomous one,
one can build analytical solutions. This method has been applied in a number
of works \cite{Belmonte-Beitia_PRL08,Yan_PRE09,Yan_PRA09,Avelar_PRE10,Cardoso_NA10,Cardoso_PLA10,Cardoso_PLA10-2,Serkin_PRA10,Serkin_JMO10,Zhang_PRA10,Yan_PLA10,He_PRE11,He_OC12,Dai_AP12,Cardoso_PRE12,Arroyo-Meza_PRE12,Cardoso_PRE13,Yomba_PLA13,He_PLA13,Zhong_O13,He_PLA14,Calaca_CNSNS14,Soloman-Raju_OC15,Temgoua_PRE15,Yang_JMP15,Kumar-De_OC15,Meza_PRE15}.
In particular, in Ref. \cite{Belmonte-Beitia_PRL08} the authors presented
localized nonlinear waves in systems with time- and space-modulated
cubic nonlinearities. More general models, including cubic-quintic
nonlinearities and modulations with dependence on both time and
space coordinates were considered in Refs. \cite{Avelar_PRE09,Zhang_PRA10,He_PRE11,Arroyo-Meza_PRE12,He_OC12,Meza_PRE15}.
Exact solutions to three-dimensional generalized NLS equations with
varying potential and nonlinearities were studied in Refs. \cite{Yan_PRE09,Yan_PRA09,Avelar_PRE10}.
Also, the modulation of breathers and rogue waves were investigated
in Refs. \cite{Avelar_PRE10,Cardoso_PLA10} and Refs. \cite{Yan_PLA10,Dai_AP12,Temgoua_PRE15},
respectively. In Refs. \cite{Cardoso_PLA10-2,Cardoso_PRE12,Yang_JMP15},
one investigated solitons of two-component systems modulated in space
and time. Moreover, in Ref. \cite{Yomba_PLA13} one studied solitons
in a generalized model, using space- and time-variable coefficients in a NLS
equation with higher-order terms. More recently, the dynamics of self-similar
waves in asymmetric twin-core fibers with Airy-Bessel modulated nonlinearity
was also investigated in \cite{Soloman-Raju_OC15}.

These previous studies have motivated the investigations of other
new possibilities, among them the quadratic-cubic nonlinear
Schr\"odinger (QCNLS) equations with inhomogeneous coefficients, in
the presence of several distinct background potentials. From the general
physical perspective, QCNLS equations have received considerable
attention in classical field theory \cite{Rajaraman_87} because
the localized solutions are non-topological or lump-like structures
\cite{Avelar_PLA09} that appear in several contexts in physics such
as q-balls, tachyon branes, and galactic dark
matter properties (see \cite{Avelar_EPJC08} and references therein).
In BEC, these equations can arise as an approximate model of a relatively
dense quasi-1D BEC with repulsive local interactions between atoms
\cite{Mateo_PRA08,Munoz-Mateo_AP09} plus a long-range dipole-dipole
attraction between them \cite{Sinha_PRL07}. Recently, the presence
of chaotic solitons in this system under nonlinearity management has
been investigated in \cite{Fujioka_CHAOS11} via numerical simulations
and variational approximation with rational and hyperbolic trial functions.
In the current work, our goal is to find analytical solutions describing localized structures 
modulated by nonautonomous QCNLS equations, allowing us to investigate different patterns of modulations.
To do this, we employ the similarity transformation technique and
direct numerical simulations to check stability of
the solutions. The problem is of current interest since we know that
information on the stability of solitonic solutions is of great significance
in the study of atomic Bose-Einstein condensates \cite{Pitaevskii_03,Kartashov_RMP11}.

The work is organized as follows. In Sec. \ref{sec:Model} we introduce
the theoretical model and apply the similarity transformation to
get information on the pattern of the inhomogeneous terms of the
QCNLS equation. Also, we present two different solutions of the autonomous QCNLS equation.
The linear stability analysis is displayed in Sec. \ref{sec:Stability}. 
Next, in the Sec. \ref{sec:Results} we consider three
different modulation patterns and show the results of the stability
tests that are obtained via direct numerical simulations. We summarize our results
and suggest new investigations in Sec. \ref{sec:Conclusion}.

%%%%%%%%%%%%%%%%%%%%%%%%%%%%%%%%%%%

\section{The quadratic-cubic model and the analytical solutions}

\label{sec:Model}

The model of interest in this work is described by the QCNLS equation
with inhomogeneous coefficients. It is given by 
\begin{equation}
i\psi_{t}=-\frac{1}{2}\psi_{xx}+V(x,t)\psi+g_{2}(t)|\psi|\psi+g_{3}(t)|\psi|^{2}\psi,\label{QCNLSE}
\end{equation}
where $\psi=\psi(x,t)$, $\psi_{t}=\partial\psi/\partial t$, $\psi_{xx}=\partial^{2}\psi/\partial x^{2}$,
$V(x,t)$ is the background or trapping potential, and $g_{2}(t)$
and $g_{3}(t)$ represent the quadratic and cubic nonlinearity intensities,
which are modulated in time, respectively. This second order partial
differential equation with quadratic and cubic time-dependent nonlinearities
is very hard to solve, but it describes very interesting physical
systems such as cigar-shaped condensates with repulsive interatomic
interactions \cite{Mateo_PRA08} plus a dipole-dipole attraction \cite{Sinha_PRL07},
i.e, Eq. (\ref{QCNLSE}) appears as an effective 1D equation that
governs the axial dynamics of mean-field cigar-shaped condensates
and accounts accurately the contribution from the transverse degrees
of freedom. Hence, our goal is to construct explicit nontrivial solutions
of this QCNLS with potentials depending on the spatial coordinate and on
time, with nonlinearities depending on time. To achieve this, we use
the following \emph{ansatz }\cite{Cardoso_PLA10} 
\begin{equation}
\psi=\rho(t)e^{i\eta(x,t)}\Phi[\zeta(x,t),\tau(t)]\label{ansatz}
\end{equation}
that connects the nonautonomous QCNLS to an autonomous QCNLS equation
\begin{equation}
i\Phi_{\tau}=-\frac{1}{2}\Phi_{\zeta\zeta}+G_{2}|\Phi|\Phi+G_{3}|\Phi|^{2}\Phi,\label{QC-const}
\end{equation}
with constant coefficients $G_{2}$ and $G_{3}$, which is easier
to solve. Note that the ansatz (\ref{ansatz}) transfer all the space and time
dependence of the coefficients of (\ref{QCNLSE}) to the external parameters, the amplitude
$\rho(t)$ and the phase $\eta(x,t)$, and now the new coordinates $\zeta(x,t)$
and $\tau(t)$ describe the space and time evolution in relation
to a frame which is moving with the localized solution.

By inserting (\ref{ansatz}) into (\ref{QCNLSE}) one gets the Eq.
(\ref{QC-const}) provided that the set of conditions 
\begin{eqnarray}
\rho_{t}+\frac{1}{2}\rho\eta_{xx} & = & 0,\label{c1}\\
\zeta_{t}+\zeta_{x}\eta_{x} & = & 0,\label{c2}\\
\tau_{t}-\zeta_{x}^{2} & = & 0,\label{c3}\\
\zeta_{xx} & = & 0.\label{c4}
\end{eqnarray}
are satisfied in order to connect the external parameters with the internal
ones. These conditions are tightly related to each
other, but they allow that we go on: for instance, from Eq. (\ref{c4}) one
derives $\zeta(x,t)=a(t)x+b(t)$,
where $a(t)$ is the inverse of the width of the localized solution
(it is positive definite) and $-b(t)/a(t)$ is the position of its
center of mass, which implies in Eq. (\ref{c3}) the following condition
$\tau(t)=\int a^{2}dt$. Then, one can obtain the amplitude and phase
of the ansatz (\ref{ansatz}), given by 
\begin{eqnarray}
\rho(t) & = & \sqrt{a},\label{amplitude}\\
\eta(x,t) & = & -\frac{a_{t}x^{2}}{2a}-\frac{b_{t}x}{a}+\epsilon(t),\label{phase}
\end{eqnarray}
respectively.

In addition, the trapping potential and nonlinear terms must have
the form 
\begin{eqnarray}
V & = & -\eta_{t}-\frac{1}{2}\eta_{x}^{2},\label{POT}\\
g_{2} & = & G_{2}\frac{\zeta_{x}^{2}}{\rho},\label{g2}\\
g_{3} & = & G_{3}\frac{\zeta_{x}^{2}}{\rho^{2}}=\frac{G_{3}g_{2}}{G_{2}\rho},\label{g3}
\end{eqnarray}
which, by using the Eqs. (\ref{amplitude}) and (\ref{phase}), can
be rewritten as 
\begin{eqnarray}
V(x,t) & = & \alpha(t)x^{2}+\beta(t)x+\delta(t),\label{V}\\
g_{2}(t) & = & G_{2}a(t)^{3/2},\label{nl2}\\
g_{3}(t) & = & G_{3}a(t),\label{nl3}
\end{eqnarray}
where 
\begin{eqnarray}
\alpha(t) & = & \left(aa_{tt}-2a_{t}^{2}\right)/2a^{2},\\
\beta(t) & = & \left(ab_{tt}-2a_{t}b_{t}\right)/a^{2},\\
\delta(t) & = & -\left(2a^{2}\epsilon_{t}+b_{t}^{2}\right)/2a^{2}.
\end{eqnarray}
Note that the modulation in the present model is completely defined
by setting the functions $a(t)$, $b(t)$, and $\epsilon(t)$. Physically,
this can be done by setting appropriately the patterns of linear and
nonlinear coefficients ($V$, $g_{2}$, and $g_{3}$, respectively)
in the system. For example, in a BEC the harmonic
potential and nonlinearities may vary in time due to the application of a modulated
laser beam that controls the interactions optically.

Next, we go further on the subject and consider an interesting possibility,
with the solution $\Phi=\varphi(\zeta)e^{-i\mu\tau}$ of Eq.~(\ref{QC-const}),
such that 
\begin{equation}
\mu\varphi=-\frac{1}{2}\varphi_{\zeta\zeta}+G_{2}\varphi^{2}+G_{3}\varphi^{3},\label{stationary}
\end{equation}
where $\mu$ is a constant and $\varphi>0$. In this case, we can get two distinct
solutions, the first one having the form 
\begin{equation}
\varphi=\frac{A}{1+B\zeta^{2}},\label{lorentizian}
\end{equation}
with $A=-4G_{2}/3G_{3}$ (assuming $A>0$), $B=-4G_{2}^{2}/9G_{3}$
and $\mu=0$. Note that we need $B>0$ for a nonsingular solution,
which implies that $G_{3}<0$ and $G_{2}>0$. The presence of a negative
$G_{3}$ implies that the system engenders focusing cubic nonlinearity,
and since $G_{2}$ is positive, one is dealing with a defocusing
quadratic nonlinearity. Also, one can obtain a different solution
\begin{equation}
\varphi=\frac{A^{\prime}}{1+B^{\prime}\cosh(\zeta)}\,,\label{cosh}
\end{equation}
where $A^{\prime}=-3/(2G_{2})$, $B^{\prime}=-\sqrt{4G_{2}^{2}-9G_{3}}/(2G_{2})$
and $\mu=-1/2$. There are two
ranges of values of $B^{\prime}$ that present nonsingular solutions,
viz., $B^{\prime}>0$ implying $G_{2}<0$ and $G_{3}<4G_{2}^{2}/9$;
$B^{\prime}<-1$ implying $G_{2}>0$ and $G_{3}<0$. Both cases are of current
interest since they can present self-focusing or self-defocusing nonlinearities (competitive or not), 
which can correspond to different types of materials constituting the nonlinear fiber or the crystal \citep{Agrawal_13}.

\section{Analysis of linear stability}

\label{sec:Stability}

To analyze the linear stability of our analytical solutions of the
autonomous QCNLS equation, we perturbed it by normal modes as 
\begin{equation}
\Phi(\zeta,\tau)  =  \left\{ \varphi(\zeta)+[v(\zeta)+w(\zeta)]e^{\lambda\tau}  
+  [v^{*}(\zeta)+w^{*}(\zeta)]e^{\lambda^{*}\tau}\right\} e^{-i\mu\tau},\label{linear_perturbation}
\end{equation}
where $v(\zeta),w(\zeta)\ll1$ are normal-mode perturbations, and
$\lambda$ is the eigenvalue of this normal mode. Inserting this perturbed
solution in (\ref{QC-const}) and linearizing, we obtain the following
linear-stability eigenvalue problem: 
\begin{equation}
\mathbf{L}\Psi=\lambda\Psi,\label{eigenvalues}
\end{equation}
where 
\begin{equation}
\mathbf{L}=i\left(\begin{array}{cc}
0 & \frac{1}{2}\nabla^{2}+\mathcal{F}_{1}\\
\frac{1}{2}\nabla^{2}+\mathcal{F}_{2} & 0
\end{array}\right),\hspace{1em}\Psi=\left(\begin{array}{c}
v\\
w
\end{array}\right),
\end{equation}
and 
\begin{eqnarray*}
\mathcal{F}_{1} & = & \mu-G_{2}\varphi-G_{3}\varphi^{2},\\
\mathcal{F}_{2} & = & \mu-2G_{2}\varphi-3G_{3}\varphi^{2},
\end{eqnarray*}
where we have assumed $\varphi$ real and positive. Here we use the
Fourier collocation method to compute eigenvalues of the linear-stability
operator $\mathbf{L}$, in which one expands the eigenfunction $\Psi$
into a Fourier series and turns Eq. (\ref{eigenvalues}) into a matrix
eigenvalue problem for the Fourier coefficients of the eigenfunction
$\Psi$. One can find examples of application of this method in Ref.
\cite{Yang_10}, and here we investigate the four new distinct possibilities
which we describe below.

%%%%%%%%%%%%%%%%%%%%%%%%%%%%%%%%%

\section{Analytical results and numerical simulations \label{sec:Results}}

We now examine the modulation of the above solutions and their stability
by numerical simulations. The numerical method is based on the $4^{\mathrm{th}}$
order split-step Crank-Nicholson algorithm in which the evolution
equation is split into several pieces (linear and nonlinear terms),
which are integrated separately. To this end, we use the steps $\Delta x=0.04$
and $\Delta t=0.001$, providing a good accuracy during the evolution
of the wave function, with fixed spatial width {[}-30,30{]}. Then,
to study stability for the above cases we employ a random perturbation
in the amplitude of the solution with the form 
\begin{equation}
\psi=\psi_{0}[1+0.05v(x)],\label{perturbed}
\end{equation}
where $\psi_{0}=\psi(x,0)$ is the analytical solution obtained via
ansatz (\ref{ansatz}) and $v\in[-0.5,0.5]$ is a real random number
with zero mean (white noise) evaluated at each point of discretization
grid in $x$-coordinate. Also, to ensure the stability of the method
we also checked the norm (power) and the energy of the solution defined
by $P=\int_{-\infty}^{\infty}|\psi|^{2}dx$ and 
\begin{equation}
E=\int_{-\infty}^{\infty}dx\left\{ \frac{1}{2}|\psi_{x}|^{2}+V|\psi|^{2}+\frac{2}{3}g_{2}|\psi|^{3}+\frac{g_{3}}{2}|\psi|^{4}\right\} ,\label{energy}
\end{equation}
respectively.

In order to focus on the practical use of the above results, in the
following we present some specific examples of typical potentials
and nonlinearities given by Eqs. (\ref{V})-(\ref{nl3}) that can be found
in experimental setups. Here, for pedagogical purpose, four distinct
cases are addressed. In the first, presented in Subsec. \ref{subsec:Vanish-potential},
we consider the system without modulation, in order to give us a landmark
about the stability of such solutions. In Subsec. \ref{subsec:Seesaw-potential}
we include a potential which is asymmetric in space and periodically modulated in
time, which is found by setting $\alpha(t)=\delta(t)=0$ and
$\beta(t)\ne0$. This type of potential is interesting because
we can see how the center of mass of the solutions behave under a
periodically oscillating uniform field. Another potential, the harmonically symmetric
in space and periodically modulated in time potential, which is obtained with $\beta(t)=\delta(t)=0$
and $\alpha(t)\ne0$, is studied in Subsec. \ref{subsec:Flying-bird-potential}.
Here the main motivation is to investigate how the solutions behave
under the effect of the squeezing and anti-squeezing produced by the
oscilatting harmonic potential. Finally, a more general case, mixing
the two previous cases, is considered in Subsec. \ref{subsec:Mixed-potential}.

Indeed, all patterns of potential and nonlinearities addressed here
are feasible in several scenarios, for example, in nonlinear fiber
optics and BECs \cite{Malomed_06,Kevrekidis_PRL03,Theis_PRL04}: the first case
can be attained by the action of a periodic heterogeneity
obtained in the fiber construction, and the second one may be driven by external potentials
and by using the Feshbach-resonance management.

\subsection{Vanishing potential \label{subsec:Vanish-potential}}

%%%%%%%%%%%%%%%%%%%%
\begin{figure}
\includegraphics[width=0.24\columnwidth]{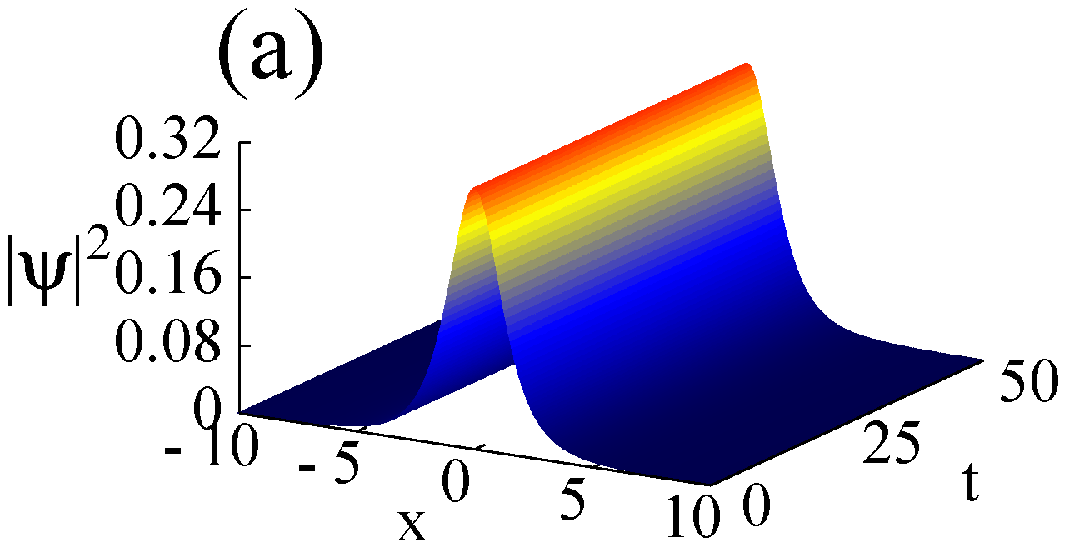} \includegraphics[width=0.24\columnwidth]{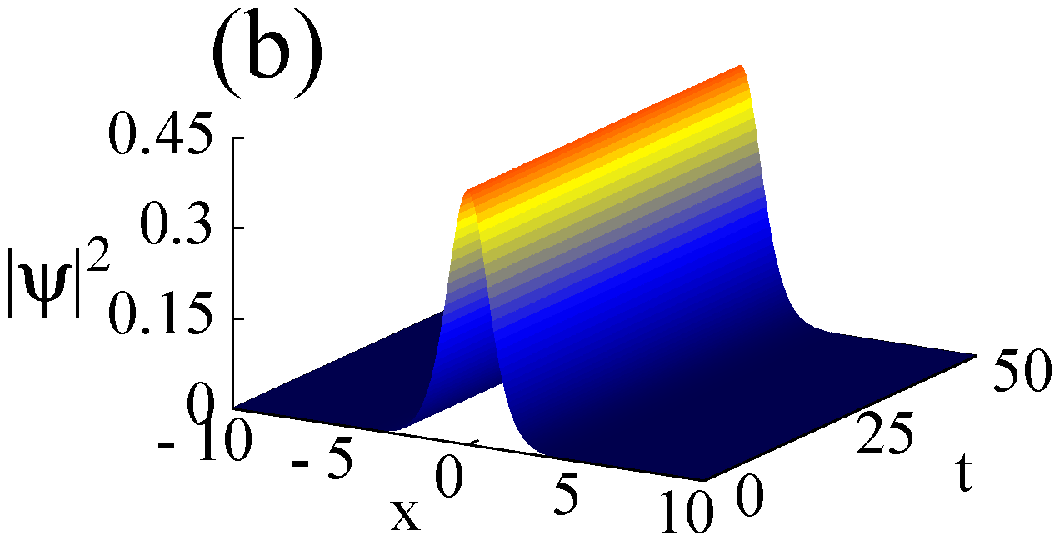}
\includegraphics[width=0.24\columnwidth]{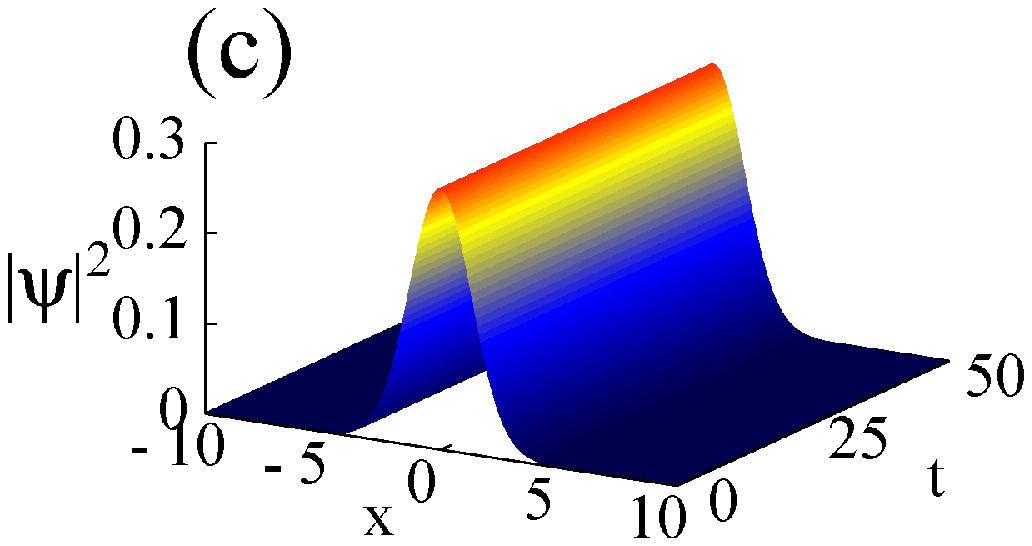} \includegraphics[width=0.24\columnwidth]{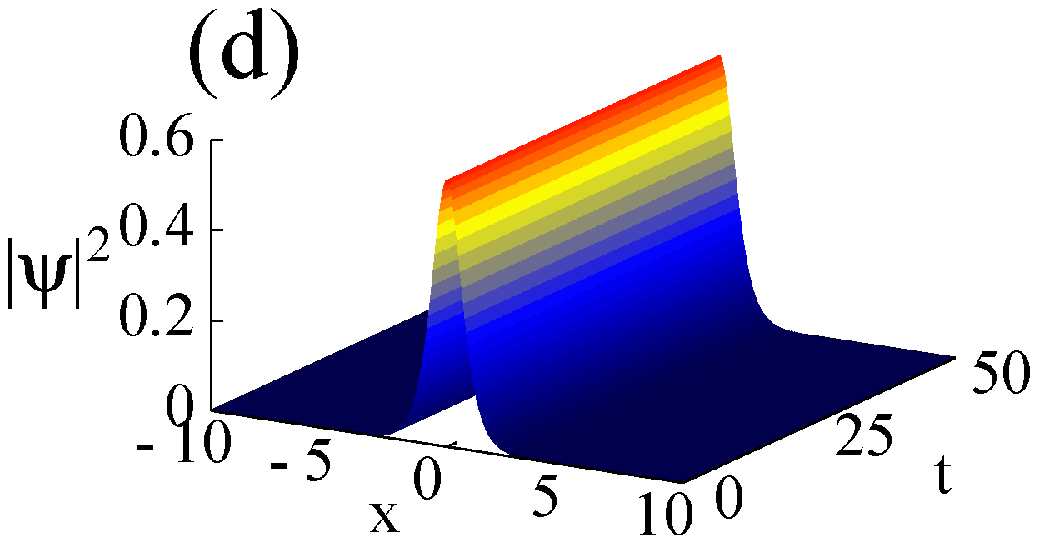}
\protect\caption{(Color online) Localized solutions $|\psi|^{2}$ obtained from the
ansatz (\ref{ansatz}) without modulation ($\alpha=\beta=\delta=0$
by setting $a=1$, $b=0$, and $\epsilon=0$). We display the analytical
results considering the static localized profiles given by Eq. (\ref{lorentizian})
in (a) and Eq. (\ref{cosh}) in (b)-(d). The values of the nonlinearities
are (a) $G_{2}=5/4$ and $G_{3}=-3$ (case A), (b) $G_{2}=-2/3$ and
$G_{3}=-1$ (case B), (c) $G_{2}=-2$ and $G_{3}=3/2$ (case C), (d)
$G_{2}=3/4$ and $G_{3}=-3$ (case D).}
\label{FN1} 
\end{figure}

%%%%%%%%%%%%%%%%%%%%%%%%%%%%%%%%%%%%%%%%%%%

In this case we assume that $V=0$ in Eq. (\ref{V}), i.e., we suppose that the system
evolves without modulation. To this end, we use $a=1$, $b=0$, and
$\epsilon=0$, for simplicity. Then, one gets $\rho=1$, $\eta=0$,
$\tau=t$, and $\zeta=x$. Also, the quadratic and cubic nonlinearities
present a constant behavior ($g_{2}=G_{2}$ and $g_{3}=G_{3}$). Then,
the solution has a constant amplitude modulation $\rho=1$ and phase
$\eta=0$. %%%%%%%%%%%%%%%%%%%%
\begin{figure}
\includegraphics[width=0.24\columnwidth]{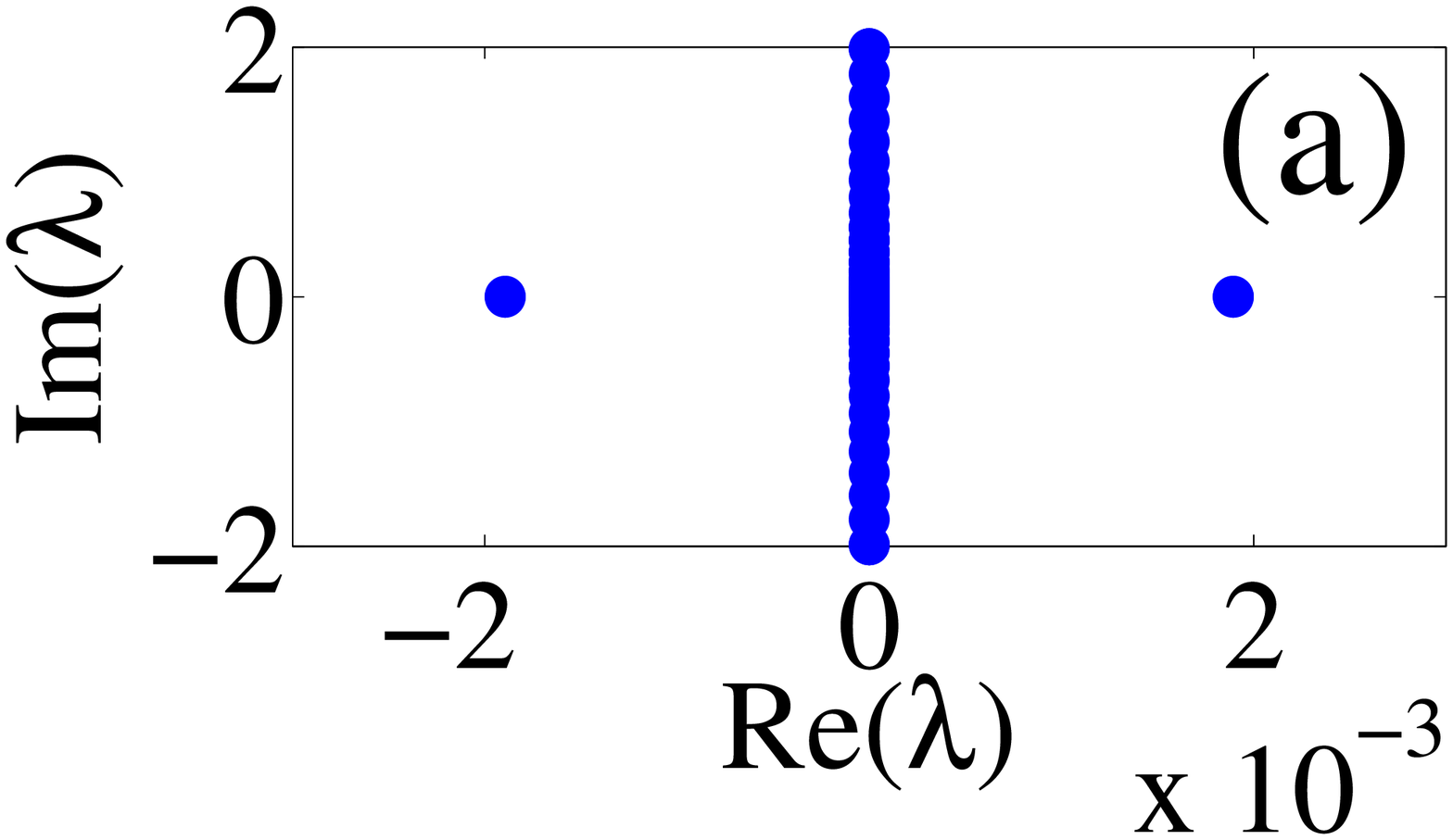} \includegraphics[width=0.24\columnwidth]{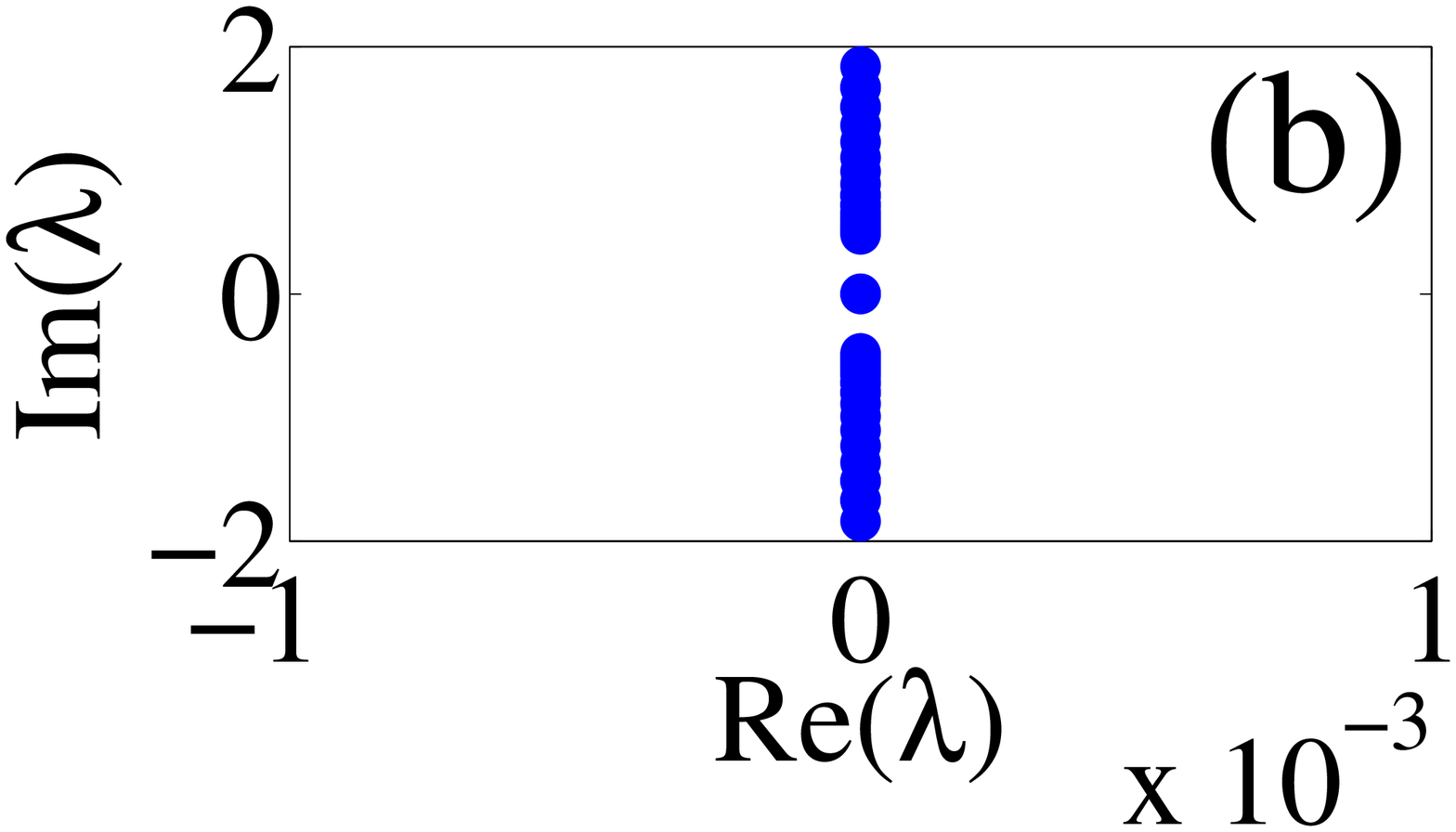}
\includegraphics[width=0.24\columnwidth]{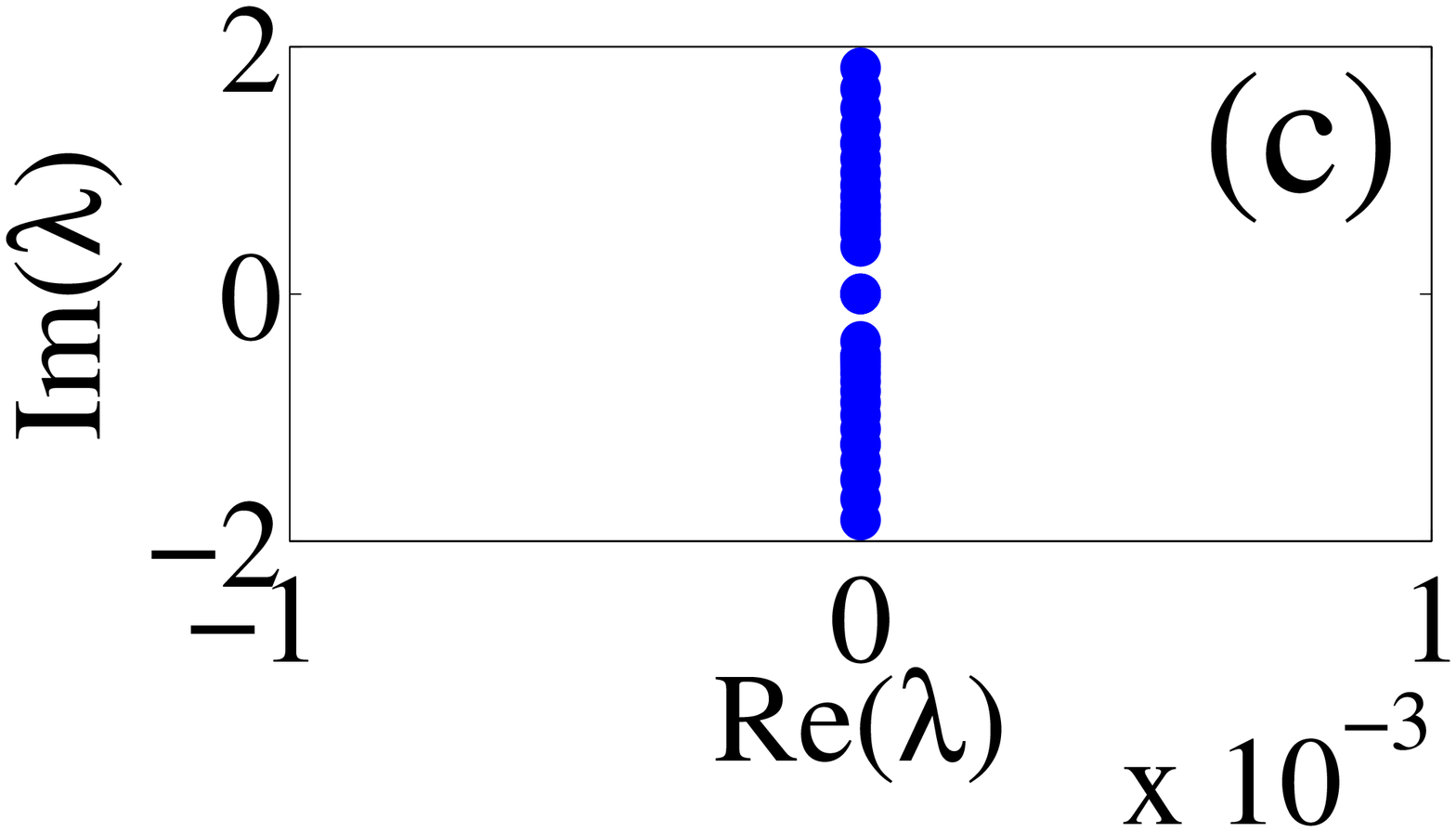} \includegraphics[width=0.24\columnwidth]{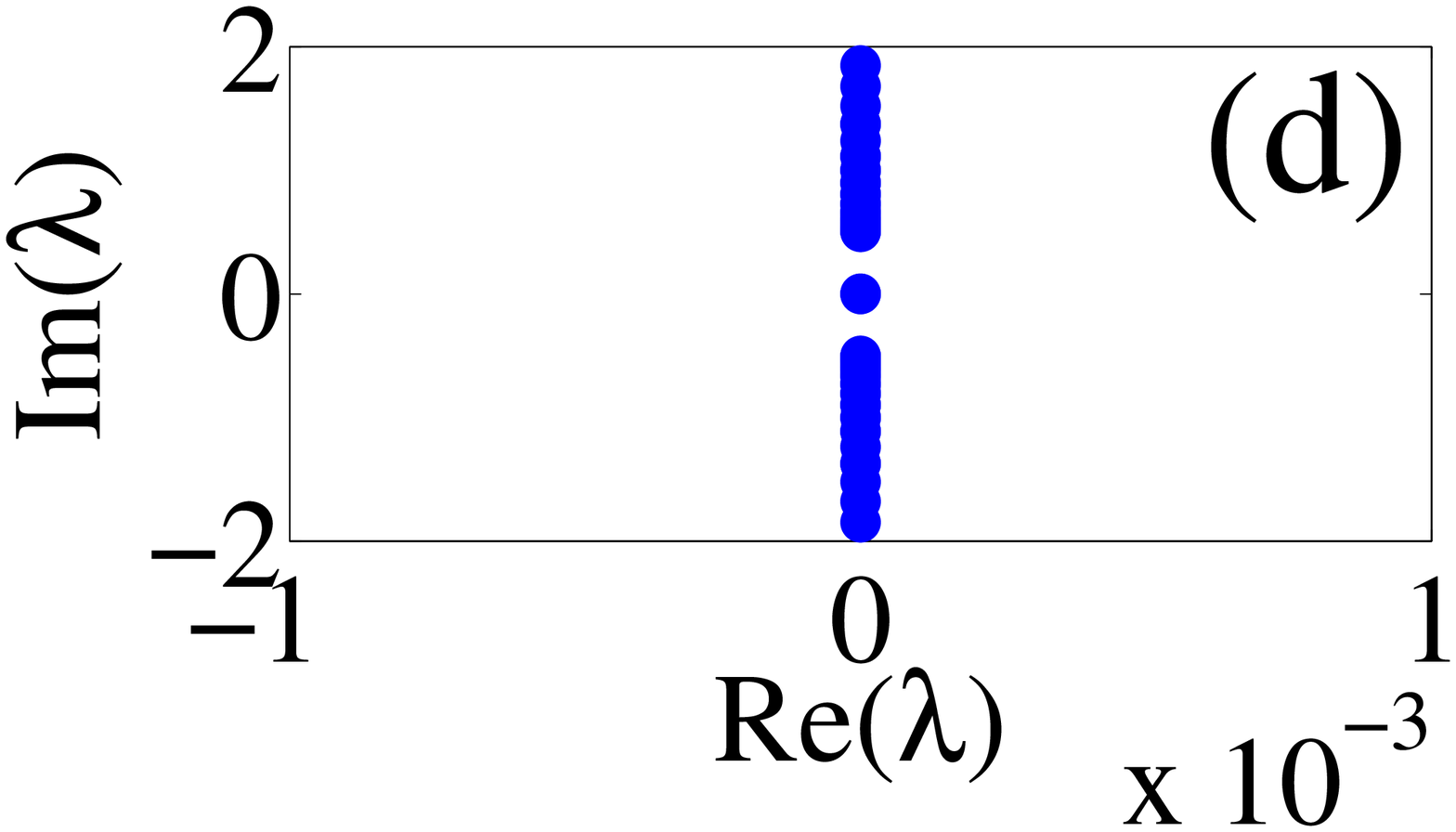}
\protect\caption{(Color online) Stability spectrum for solitary waves shown in Fig.
\ref{FN1} obtained via the linear-stability eigenvalue problem (\ref{eigenvalues}).
The parameters used here are the same of Fig. \ref{FN1}.}
\label{FN2} 
\end{figure}

%%%%%%%%%%%%%%%%%%%%%%%%%%%%%%%

In Figs. \ref{FN1}(a)-\ref{FN1}(d) we show the profiles of the localized
solutions $|\psi|^{2}$ for the cases given by Eqs. (\ref{lorentizian})
and (\ref{cosh}), in the absence of modulation (The values chosen
by us for the nonlinearities are such that the norm of the solution
approaches $1$). Note that we choose three ranges of values for the
nonlinearities, namely, $g_{2}<0$ and $g_{3}<0$ (both self-focusing
in Fig.~\ref{FN1}(b)), $g_{2}<0$ and $g_{3}>0$ (competing type-1
in Fig.~\ref{FN1}(c)), and $g_{2}>0$ and $g_{3}<0$ (competing
type-2 in Figs.~\ref{FN1}(a) and \ref{FN1}(d)). Also, we show
in Figs.~\ref{FN2}(a)-\ref{FN2}(d) the linear stability analysis
corresponding to the cases presented in Figs.\ref{FN1}(a)-\ref{FN1}(d),
where we display the real and imaginary parts of the eigenvalue given
by Eq. (\ref{eigenvalues}). Note that if $\mathrm{Re}(\lambda)\neq0$,
one gets a linearly unstable solution (cf. Eq.~(\ref{linear_perturbation})).
In our example, the Lorentzian-type solution (Eq.~(\ref{lorentizian}))
is prone to be unstable while the others solutions are linearly stable.
From now on, we will call the examples for those specific choices
of nonlinearities presented in Figs.~\ref{FN1}(a)-(d) by cases A,
B, C, and D, respectively.

%%%%%%%%%%%%%%%%%%%%%%%%%%%%%%%%

\subsection{Seesaw potential\label{subsec:Seesaw-potential}}

We now analyze a potential with linear modulation in $x$-coordinate
and periodic modulation in $t$. So, we choose $a=1$ and $b=-\sin(\omega t)$,
such that, $\alpha=0$, $\beta=\omega^{2}\sin(\omega t)$, and $\delta=0$
with a suitable adjustment of the function $\epsilon(t)$. In this
case, the amplitude and phase of the solution will be given by $\rho=1$
and $\eta=\omega x\cos(\omega t)-\frac{1}{4}\omega[\cos(\omega t)\sin(\omega t)+\omega t]$,
respectively. Also, one gets $\zeta=x-\sin(\omega t)$, $\tau=t$,
$g_{2}=G_{2}$ and $g_{3}=G_{3}$ (constant nonlinearities), and a
seesaw potential with the form 
\[
V=\omega^{2}x\sin(\omega t).
\]
Note that the amplitude of the above potential depends on the square
of oscillation frequency of the temporal modulation.

\begin{figure}
\includegraphics[width=0.24\columnwidth]{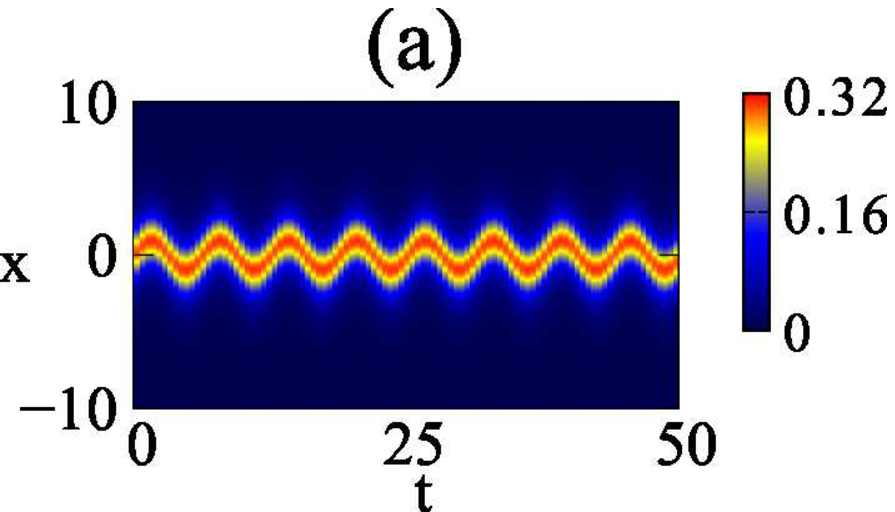} \includegraphics[width=0.24\columnwidth]{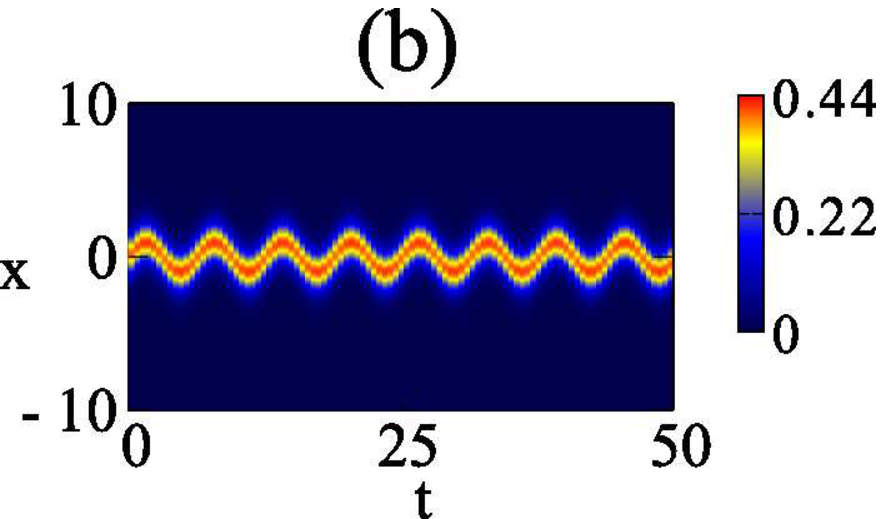}
\includegraphics[width=0.24\columnwidth]{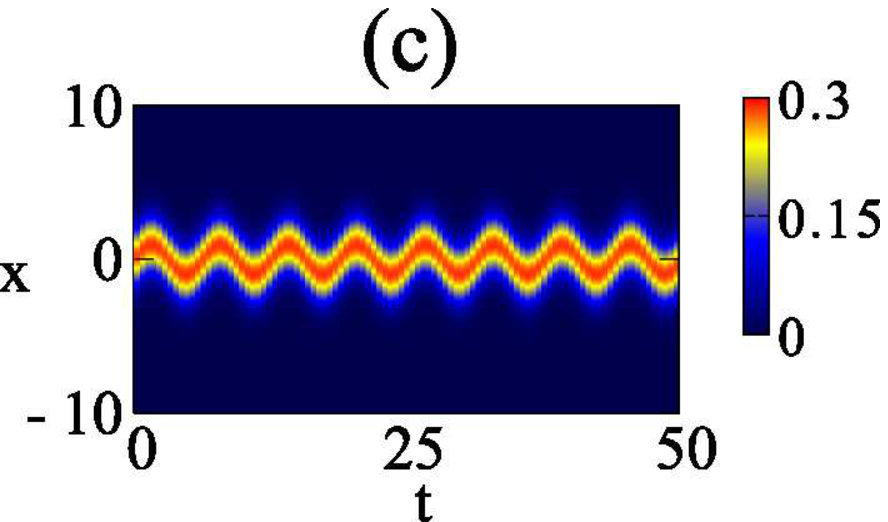} \includegraphics[width=0.24\columnwidth]{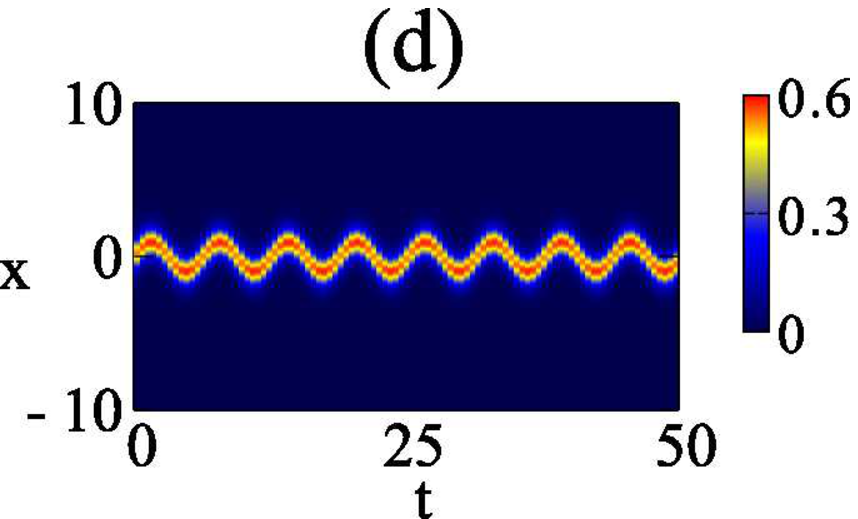}
\protect\caption{(Color online) Localized solutions $|\psi|^{2}$ obtained from the
ansatz (\ref{ansatz}) considering $\alpha=\delta=0$ and $\beta=\omega^{2}\sin(\omega t)$
(here with $\omega=1$, for simplicity). The profiles shown in (a)-(d)
corresponds to those non-modulated cases presented in Figs. \ref{FN1}(a)-\ref{FN1}(d),
respectively, but now with modulation. The values of the nonlinearities
are the same used in Fig.~\ref{FN1}.}
\label{FN3} 
\end{figure}

We display in Fig. \ref{FN3} the analytical profiles ($|\psi|^{2}$)
of the localized solutions modulated by the seesaw potential. Note
that in Figs. \ref{FN3}(a)-\ref{FN3}(d) we contemplate the same
cases shown in Figs. \ref{FN1}(a)-\ref{FN1}(d), respectively, now
with modulation of a seesaw potential. We stress that in the present
case, the linear stability analysis employed in the previous case
do not work anymore. Then, we analyze the stability of the solutions
by direct numerical simulations of the perturbed profiles (Eq. (\ref{perturbed})).
Based on the analytical solutions, we expect stable solutions when
the variance in $x$ is approximately ``constant'', i.e, $var(x)=\langle x^{2}\rangle-\langle x\rangle^{2}$,
with $\langle\bullet\rangle=\int_{-\infty}^{\infty}\bullet|\psi(x,t)|^{2}dx$.
Indeed, due to the perturbations it will only suffer small random
variations. 

\begin{figure}
\centering \includegraphics[width=0.24\columnwidth]{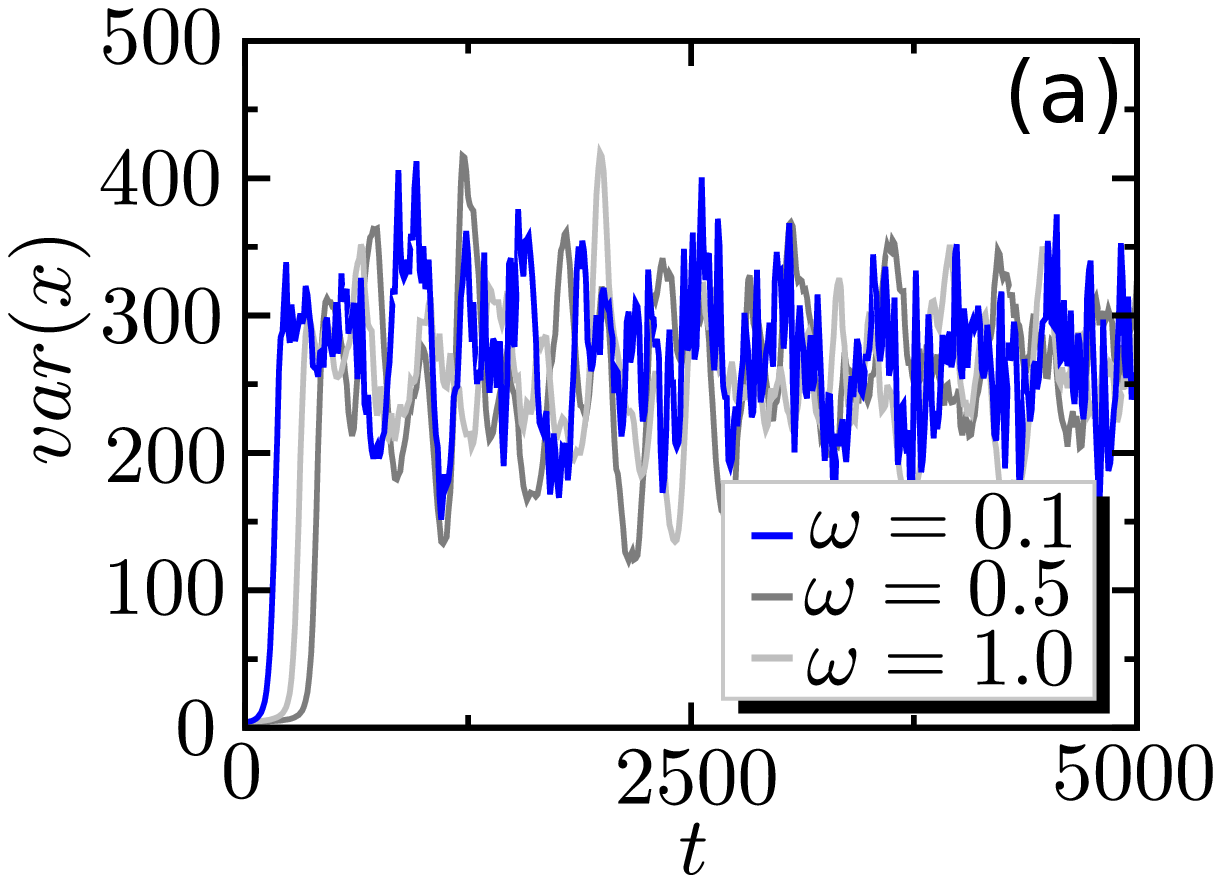} \includegraphics[width=0.24\columnwidth]{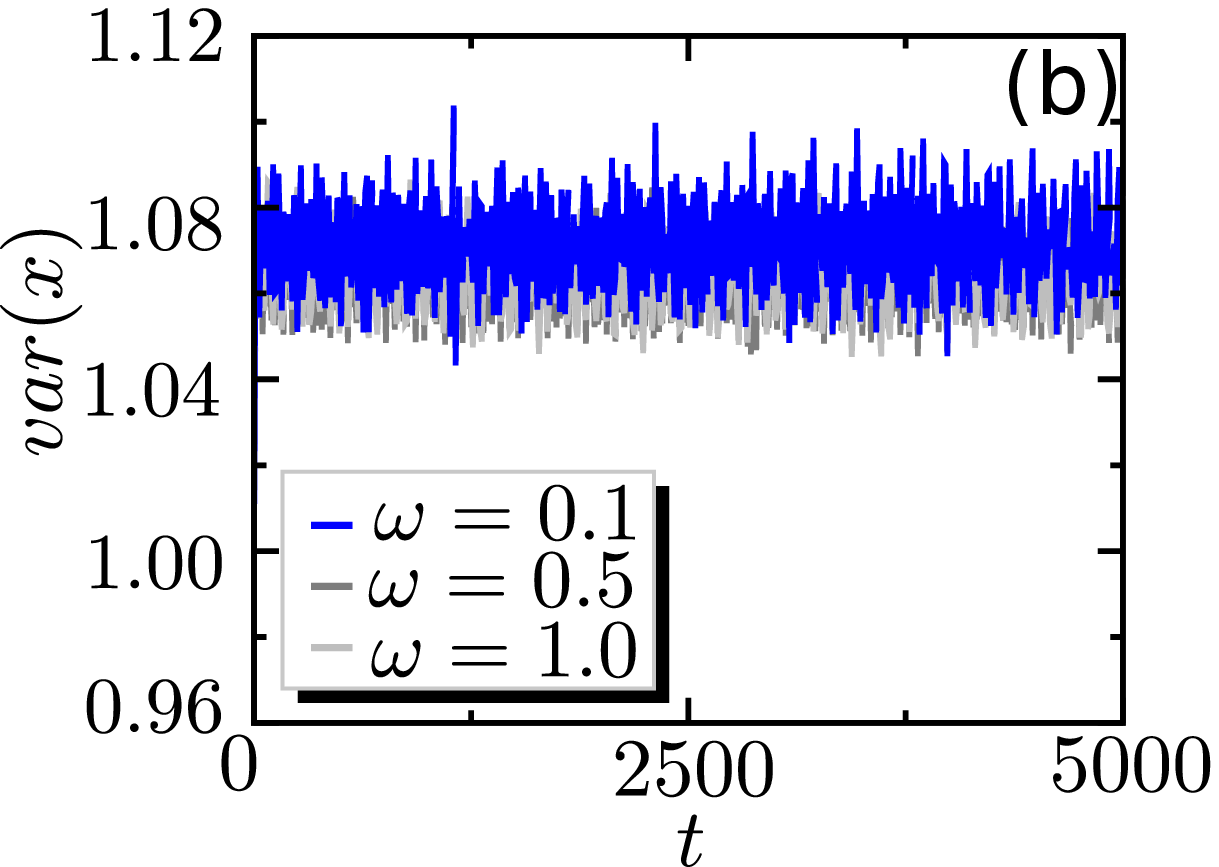}
\includegraphics[width=0.24\columnwidth]{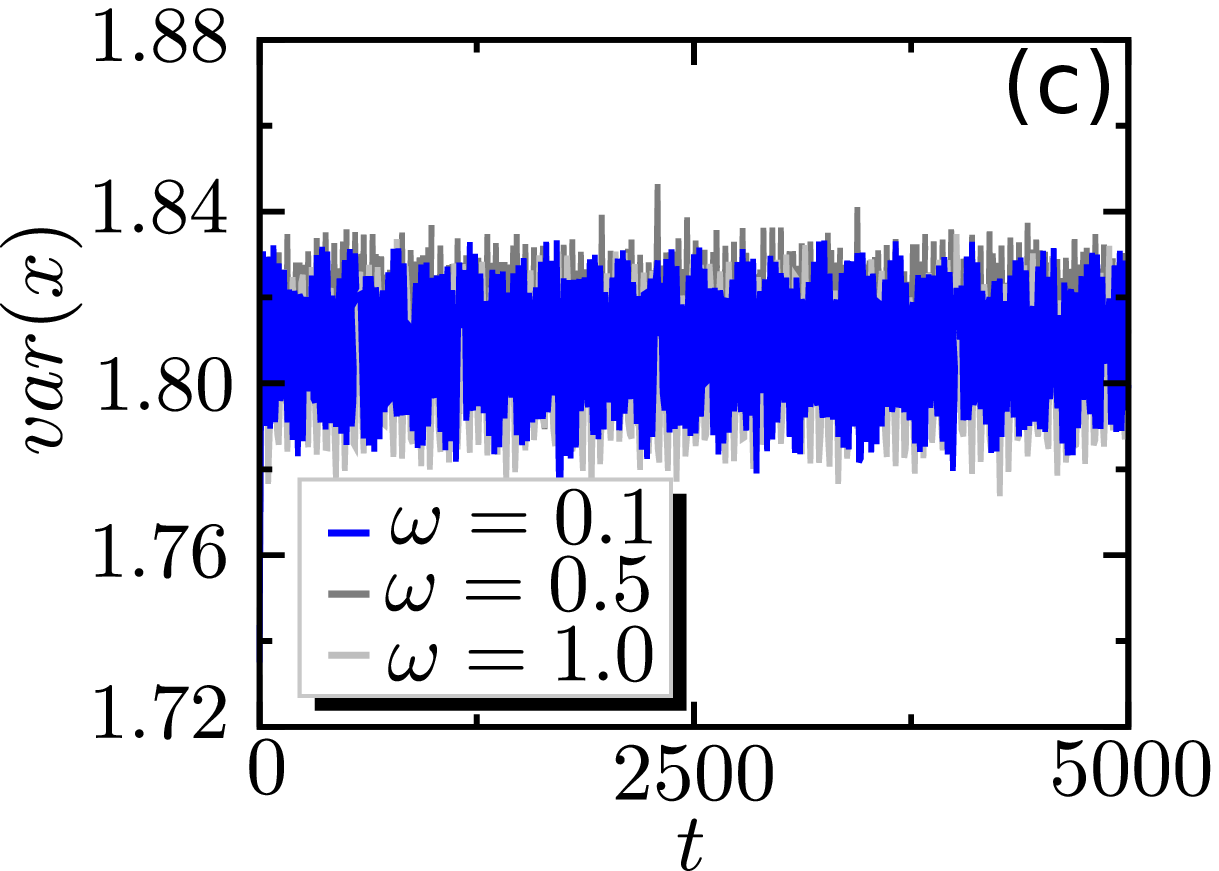} \includegraphics[width=0.24\columnwidth]{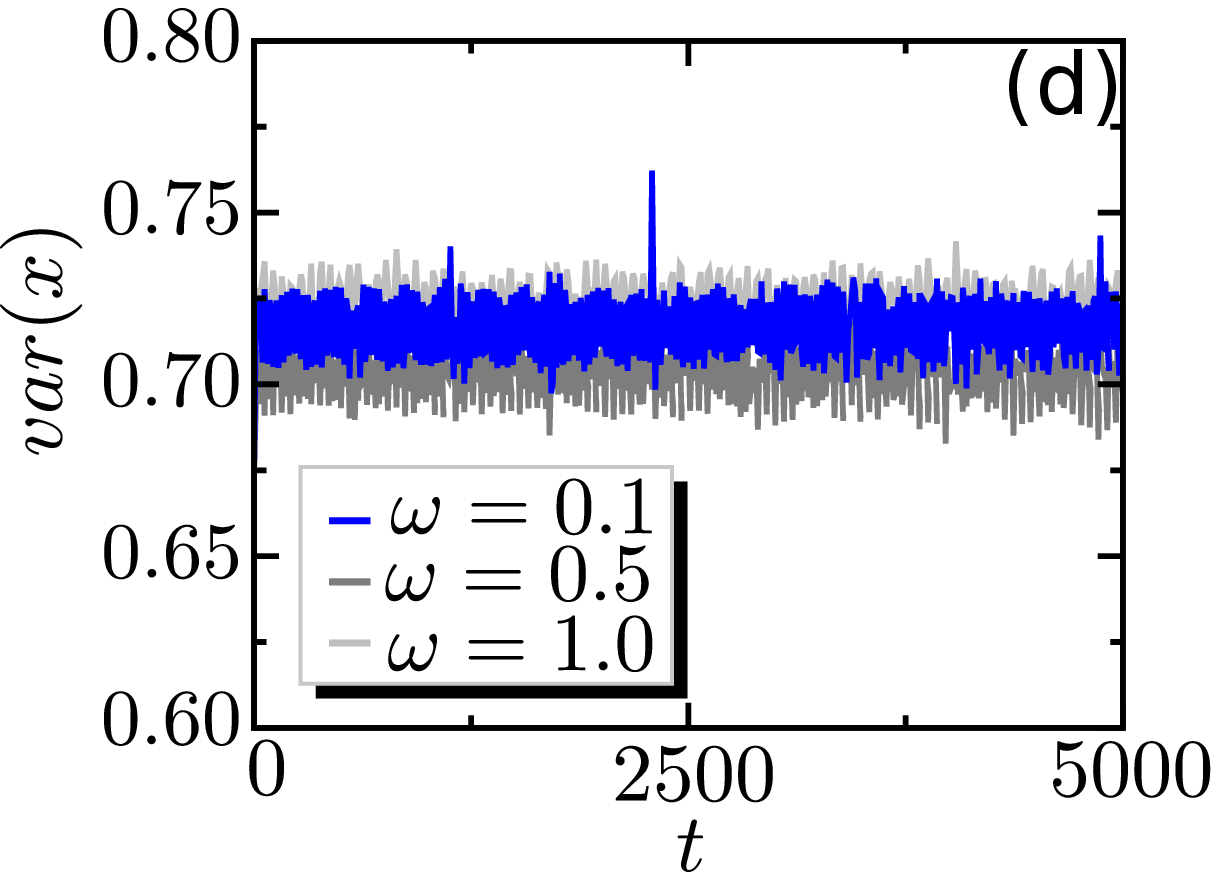}

\caption{(Color online) Stability tests via direct numerical simulations, using
the input state given by (\ref{perturbed}) with different modulation
frequencies, \emph{viz.} $\omega=0.1$, $0.5$, and $1.0$. We show
in panels (a)-(d) the variance of $x$ versus $t$ ($var(x)$)
corresponding to the cases displayed in Figs. \ref{FN3}(a)-\ref{FN3}(d),
respectively.}
\label{FN4} 
\end{figure}

In Figs. \ref{FN4}(a)-\ref{FN4}(d) we display the variance of $x$
versus $t$. Note that we do not present the 3D profiles to avoid
problems of graphical resolution because the number of oscillations
up to $t=5000$, but they were taken into account everywhere.
As a conclusion, the result of Fig. \ref{FN4}(a) shows that the unstable
solution (one whose instability is shown in Fig. \ref{FN2}(a)) remains
unstable under the present modulation. Also, those stable solutions,
whose stabilities are shown in Figs. \ref{FN2}(b)-\ref{FN2}(d),
remain stable. We stress that all results were verified for different
values of $\omega$, varying by the step $0.05$ into the range $[0,1.0]$.

%%%%%%%%%%%%%%%%%

%%%%%%%%%%%%%%%%%%%%%%%%%%%%%%%%%%%%%%%%%%%%%%%%%%%%%%%%%%%

%%%%%%%%%%%%%%%%%%%%%%%%%%%%%

\subsection{Flying-bird potential\label{subsec:Flying-bird-potential}}

Here we assume a quadratic modulation in $x$-coordinate with a periodic
modulation in $t$-coordinate. Thus, we use $a=1+\gamma\cos(\omega t)$
(with $\gamma<1$) and $b=\epsilon=0$, getting 
\begin{equation}
\alpha=\frac{\gamma\omega^{2}(\gamma\cos^{2}(\omega t)-\cos(\omega t)-2\gamma)}{2(\gamma\cos(\omega t)+1)^{2}},\label{alpha_C}
\end{equation}
$\beta=0$, and $\delta=0$. So, the amplitude and phase of the solution,
the external potential and the modulated nonlinearities will be given
by $\rho=\sqrt{1+\gamma\cos(\omega t)}$, $\eta=\gamma\omega x\sin(\omega t)/\{2[\gamma\cos(\omega t)+1]\}$,
$V=\alpha(t)x^{2}$ (with $\alpha$ given by Eq.~(\ref{alpha_C})),
$g_{2}=G_{2}[1+\gamma\cos(\omega t)]^{3/2}$, and $g_{3}=G_{3}[1+\gamma\cos(\omega t)]$,
respectively.

\begin{figure}
\includegraphics[width=0.24\columnwidth]{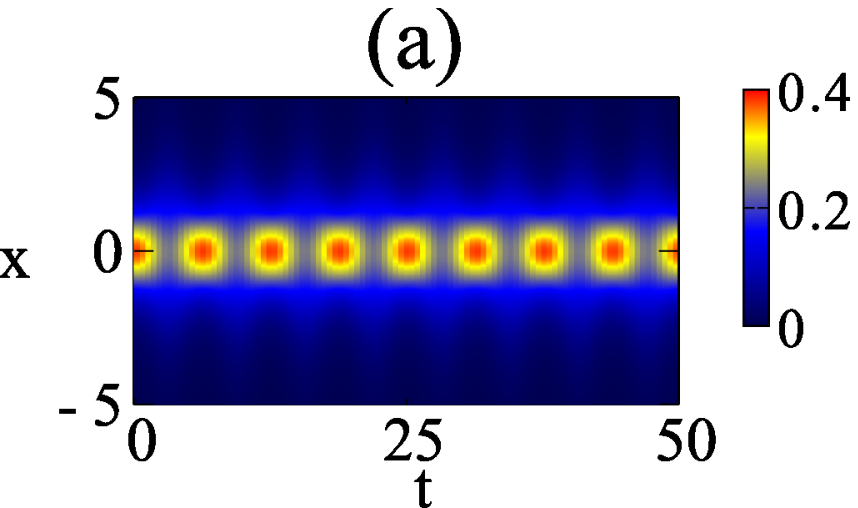} \includegraphics[width=0.24\columnwidth]{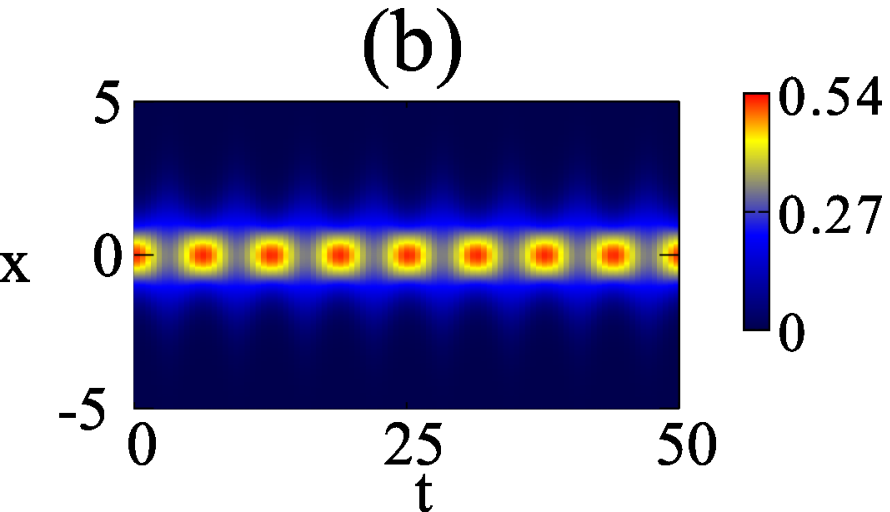}
\includegraphics[width=0.24\columnwidth]{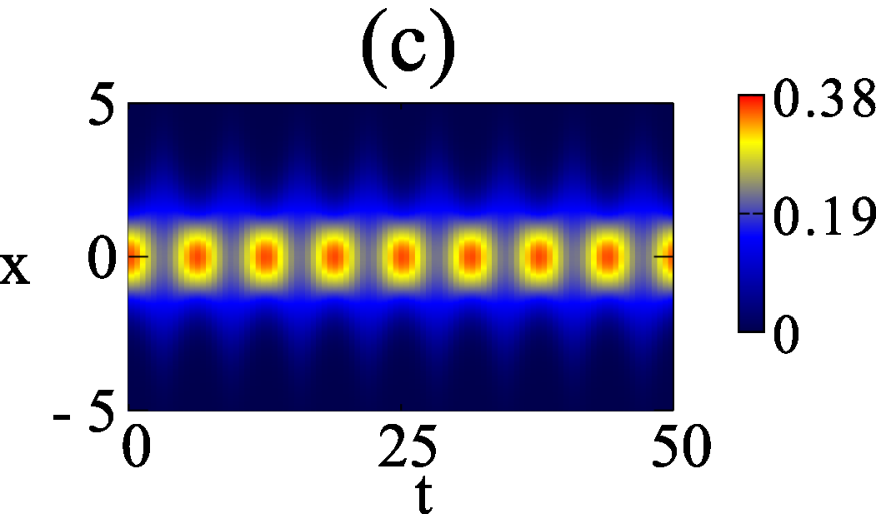} \includegraphics[width=0.24\columnwidth]{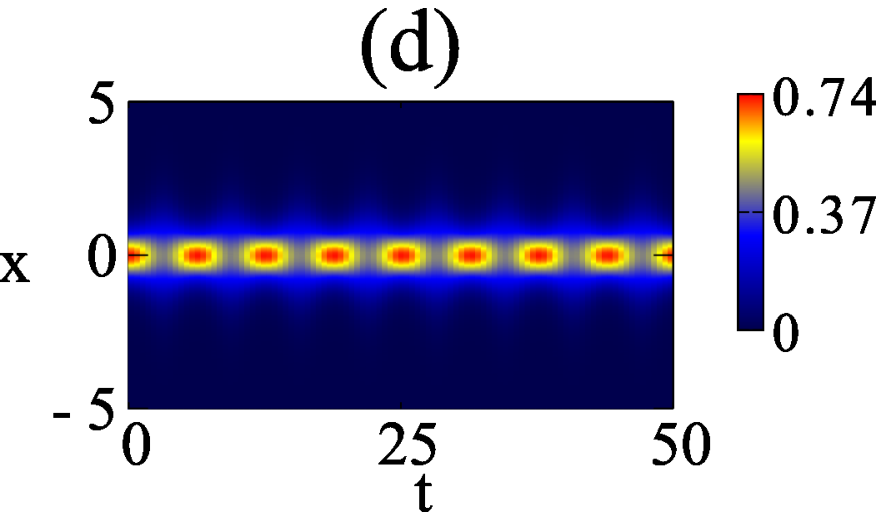}
\protect\caption{(Color online) Localized solutions $|\psi|^{2}$ obtained from the
ansatz (\ref{ansatz}) considering $a=1+\gamma\cos(\omega t)$, $b=0$,
and $\epsilon=0$. The profiles shown in (a)-(d) corresponds to the
non-modulated cases presented in Figs. \ref{FN1}(a)-\ref{FN1}(d),
respectively, but now with modulation. The values of the nonlinearities
are the same used in Fig. \ref{FN1} plus $\omega=1$ and $\gamma=1/4$.}
\label{FN5} 
\end{figure}

\begin{figure}
\centering \includegraphics[width=0.24\columnwidth]{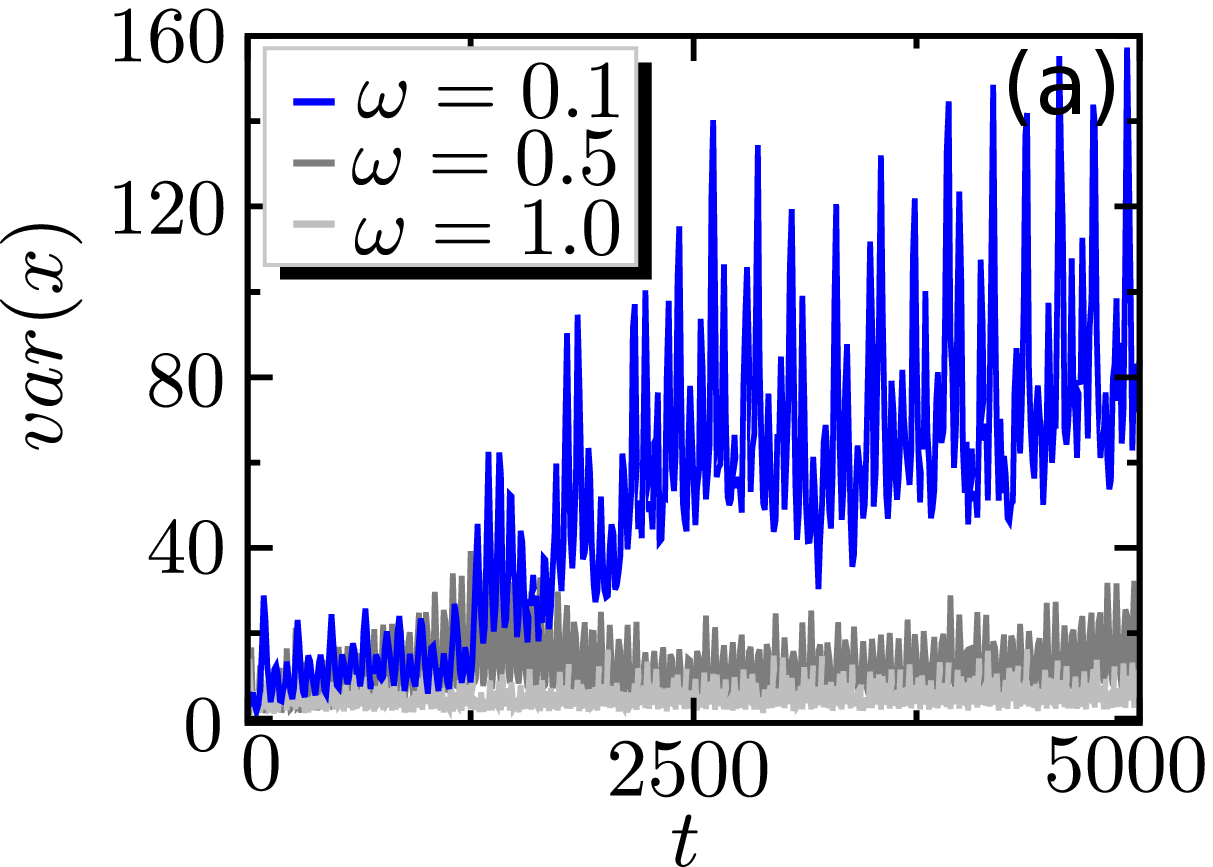} \includegraphics[width=0.24\columnwidth]{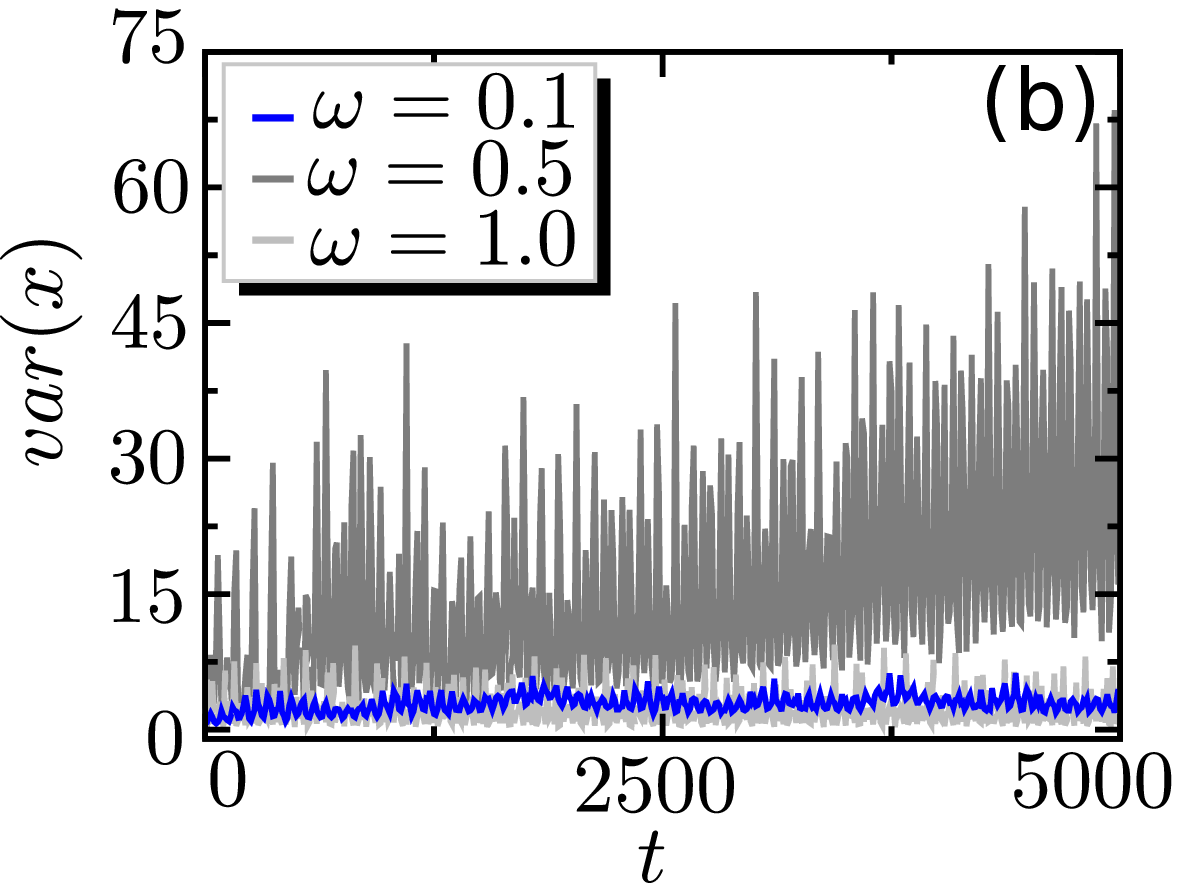}
\includegraphics[width=0.24\columnwidth]{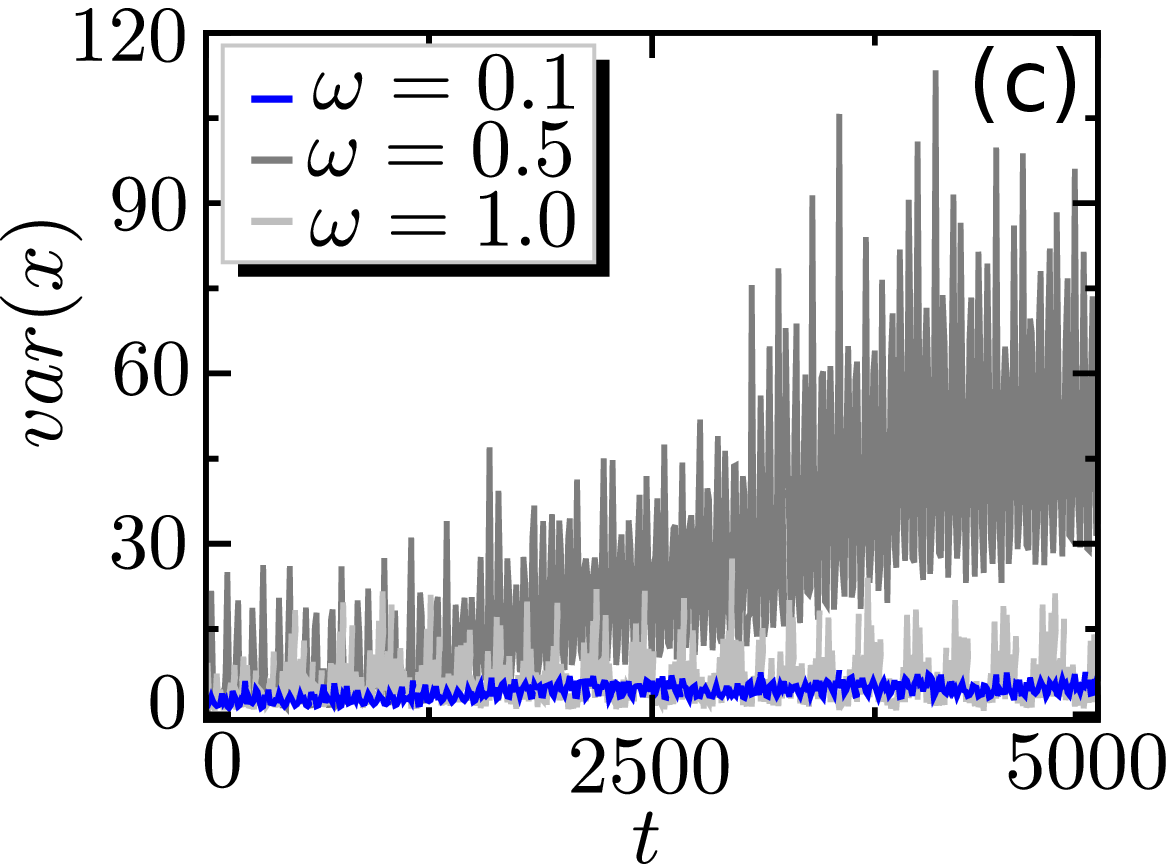} \includegraphics[width=0.24\columnwidth]{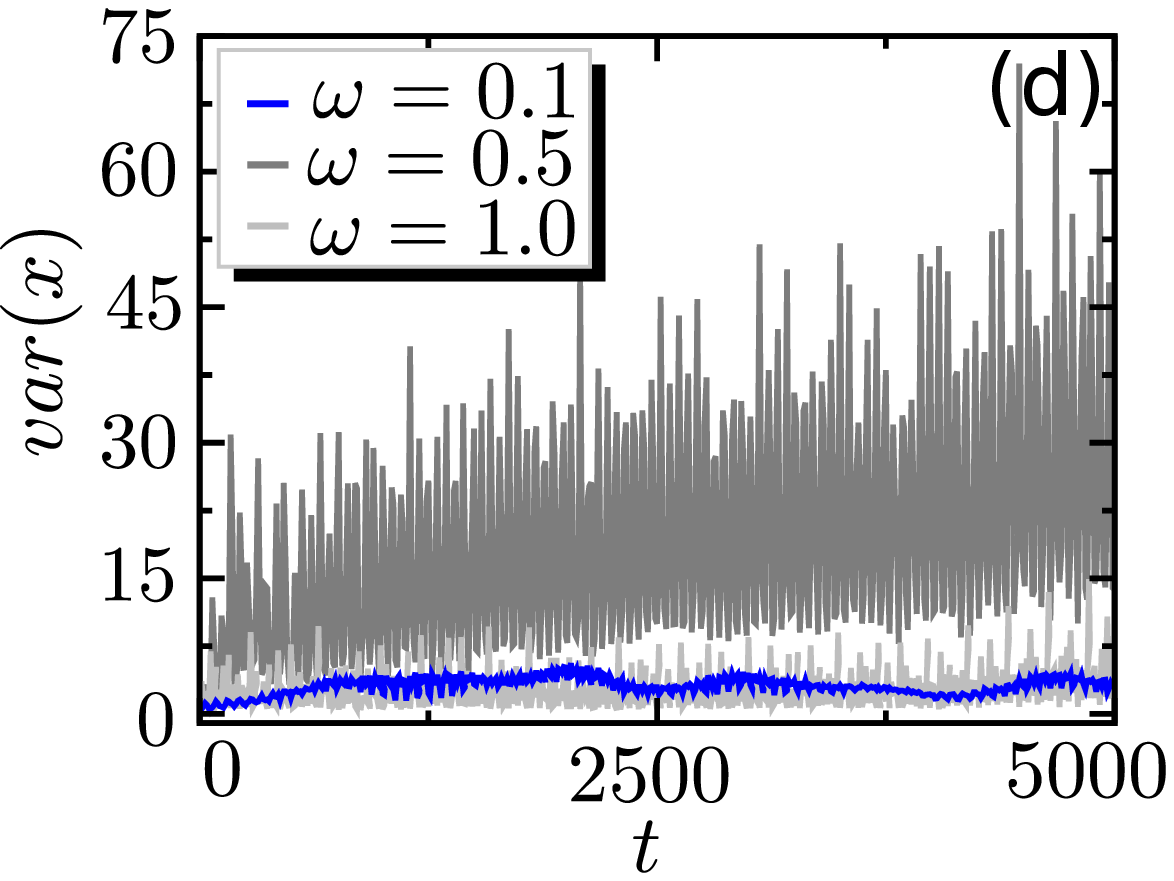}
\caption{(Color online) Stability tests via direct numerical simulations, using
the input state given by (\ref{perturbed}) with different modulation
frequencies, \emph{viz.} $\omega=0.1$, $0.5$, and $1.0$. We show
in panels (a)-(d) the variance of $x$ versus $t$ ($var(x)$)
corresponding to the cases displayed in Figs. \ref{FN5}(a)-\ref{FN5}(d),
respectively.}
\label{FN6} 
\end{figure}

Analytical profiles of the modulated solutions are shown in Fig.~\ref{FN5}.
Note that a breathing pattern is obtained since we have nonlinearities
varying harmonically while the potential presents an attractive-to-expulsive
harmonic change in its profile. In Figs. \ref{FN6}(a)-(d) we show
the time evolution of the variance of $x$ obtained by direct numerical
simulations of Eq. (\ref{QCNLSE}). Now, differently from the variance
predicted for the seesaw potential (Subsec. \ref{subsec:Seesaw-potential}),
here this parameter will oscillate around a constant value, which
reflects the breathing pattern of the solutions. 

We found different regions of stability/instability for each case.
Interestingly, we observe in the case A that the Lorentzian solution
becomes stable for $\omega\in[0.2,1.0]$. This behavior is observed
in the results shown in Fig. \ref{FN6}(a), where one can see that
the variance of the curve for $\omega=0.1$ increases in an unpredictable
fashion while the other ones remain oscillating around a constant
value. In the case B (Fig. \ref{FN6}(a)) we found an unstable region
for $0.4\leq\omega\leq0.55$ and the solutions remain stable outside
this region. Differently form the case B, in the cases C and D (Figs.
\ref{FN6}(c) and \ref{FN6}(d)) we found two unstable regions: $\omega\in[0.3,0.55]$
and $\omega=0.95$ for the case C and $\omega\in[0.25,0.65]$ and
$\omega=0.75$ for the case D. Note that in the Figs. \ref{FN6}(b)-\ref{FN6}(d)
the variances for $\omega=0.5$ present a signature of this instability.

%%%%%%%%%%%%%%%%%%%%%

%%%%%%%%%%%%%%%%%%%%%%%%%%%%%%%%%%%%%%%%%%%%%%%%%%%

%%%%%%%%%%%%%%%%%%%%%%%%%%%

%%%%%%%%%%%%%%%%%%%%%%%%%%%%%%%%%%%%%%

\subsection{Mixed potential\label{subsec:Mixed-potential}}

In this case, we consider a mixed potential with the form $V=\alpha(t)x^{2}+\beta(t)x$,
and we take $a=1+\gamma\cos(\omega t)$ and $b=-\sin(\omega t)$.
We choose $\epsilon$ in a way such that $\delta=0$. The temporal
modulation functions for the potential will then be written in the
form: 
\begin{equation}
\beta=\frac{\omega^{2}\sin(\omega t)[1-\gamma\cos(\omega t)]}{[1+\gamma\cos(\omega t)]^{2}},\label{beta_MP}
\end{equation}
with $\alpha$ given by Eq. (\ref{alpha_C}). With the above choices, it is possible to check that
$\rho=\sqrt{1+\gamma\cos(\omega t)}$ and $\eta=\omega x[\gamma x\sin(\omega t)+2\cos(\omega t)]/[2\gamma\cos(\omega t)+2]+\epsilon(t)$
for the amplitude and phase of the solution, respectively, and that
$g_{2}=G_{2}[1+\gamma\cos(\omega t)]^{3/2}$ and $g_{3}=G_{3}[1+\gamma\cos(\omega t)]$.
The modulated coordinates take the forms $\zeta=[1+\gamma\cos(\omega t)]x-\sin(\omega t)$
and $\tau=\{\gamma[4+\gamma\cos(\omega t)]\sin(\omega t)+\omega t(2+\gamma^{2})\}/2\omega$.
%%%%%%%%%%%%%%%%%%%%%%%
\begin{figure}
\includegraphics[width=0.24\columnwidth]{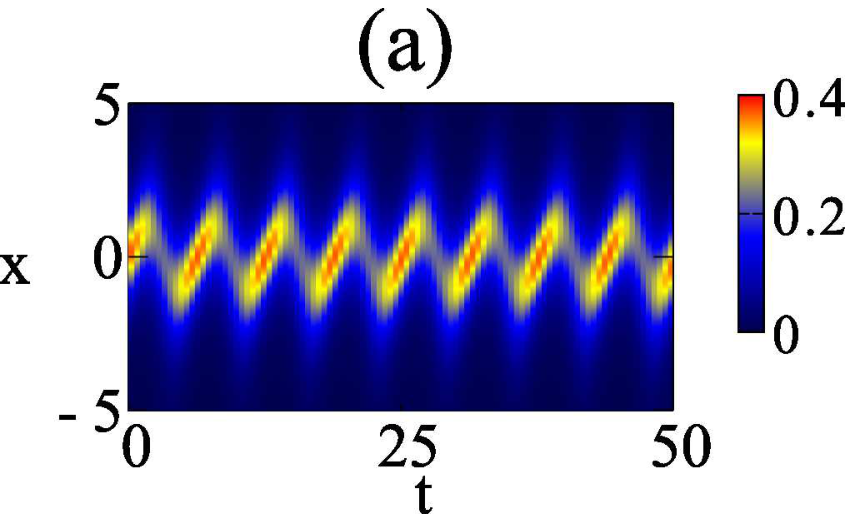} \includegraphics[width=0.24\columnwidth]{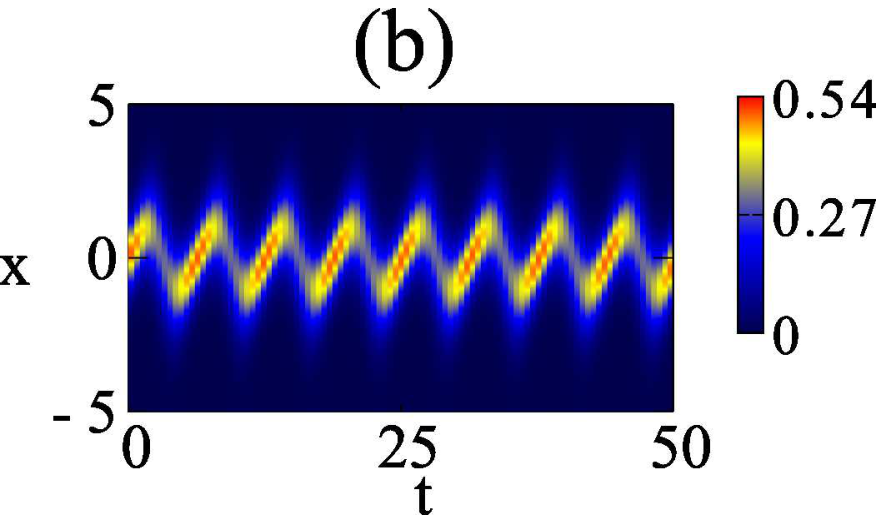}
\includegraphics[width=0.24\columnwidth]{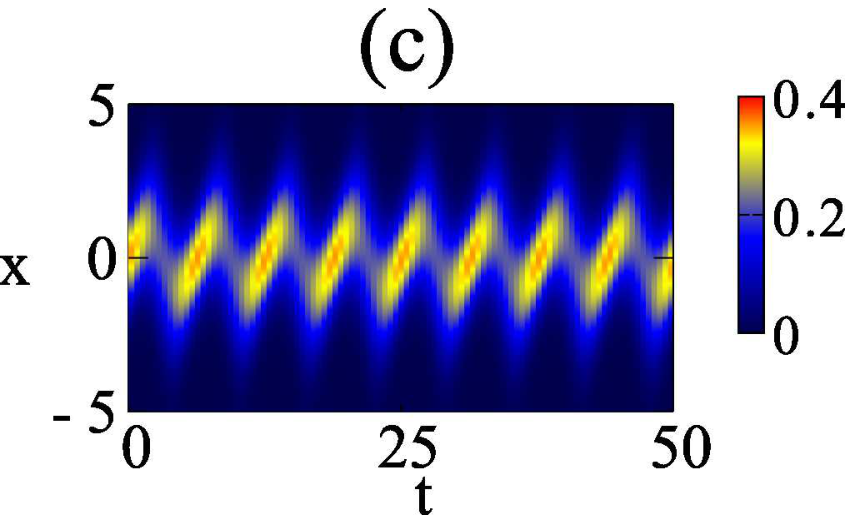} \includegraphics[width=0.24\columnwidth]{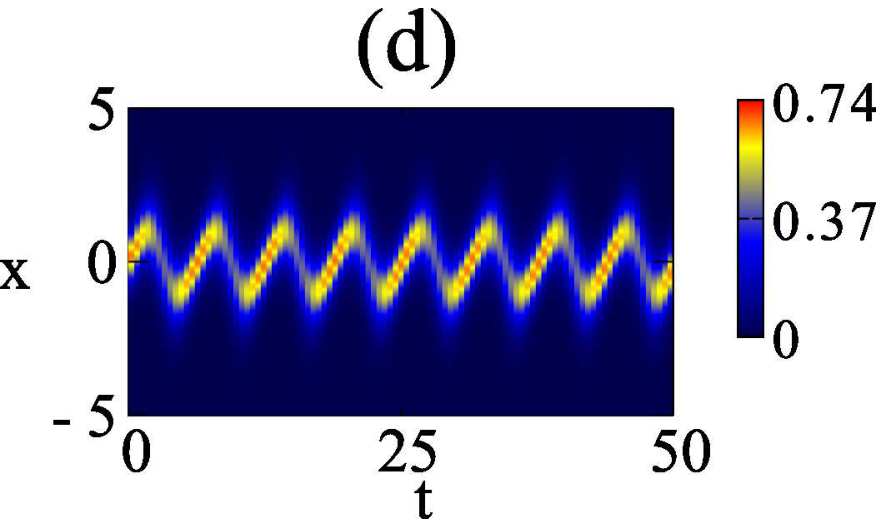}
\protect\caption{(Color online) Localized solutions $|\psi|^{2}$ obtained from the
ansatz (\ref{ansatz}) considering $a=1+\gamma\cos(\omega t)$, $b=-\sin(\omega t)$,
and $\epsilon$ in a such way that allow us to get $\delta=0$. The
profiles shown in (a)-(d) corresponds to the non-modulated cases presented
in Figs. \ref{FN1}(a)-\ref{FN1}(d), respectively, but now with modulation.
The values of the nonlinearities are the same used in Fig. \ref{FN1}
plus $\omega=1$ and $\gamma=1/4$.}
\label{FN7} 
\end{figure}

%%%%%%%%%%%%%%%%%%%%%%%%

%%%%%%%%%%%%%%%%%%%%%%%%%%%%
\begin{figure}[tb]
\centering \includegraphics[width=0.24\columnwidth]{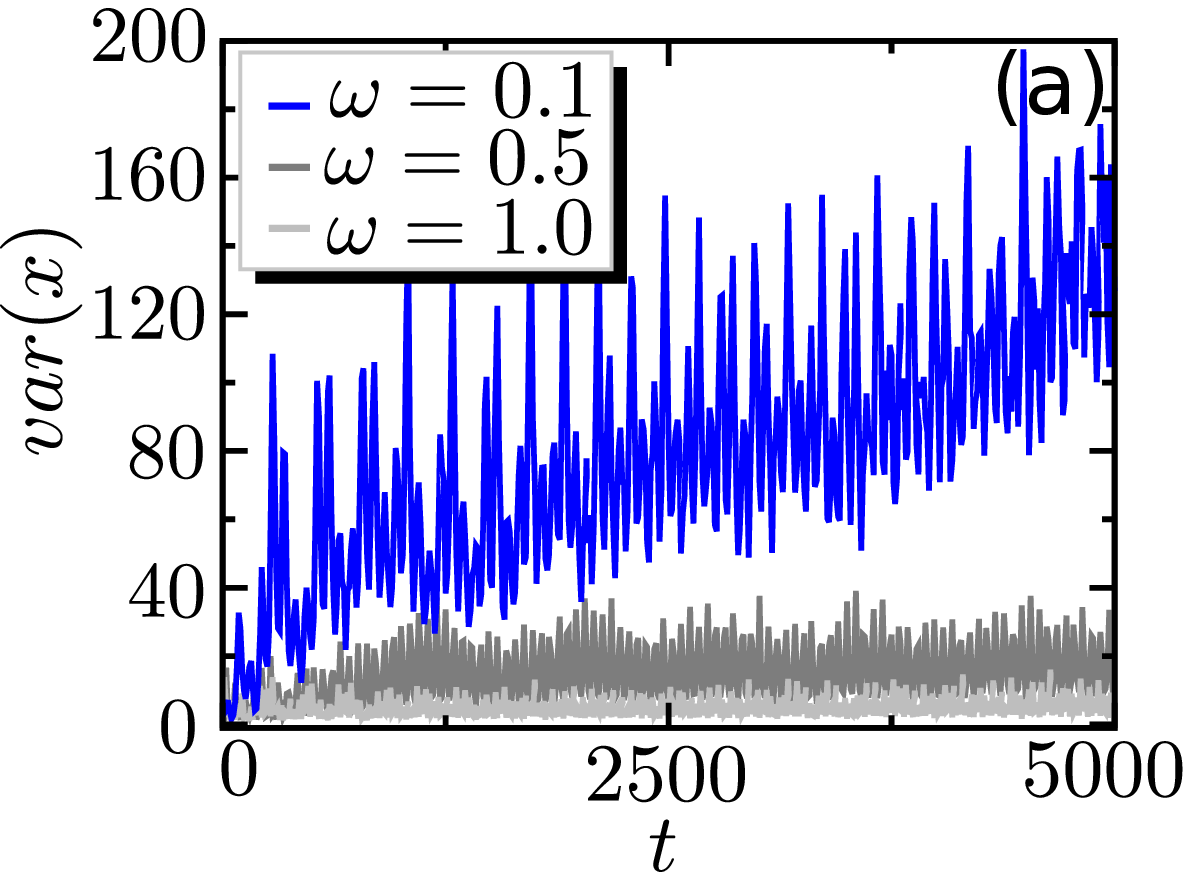} \includegraphics[width=0.24\columnwidth]{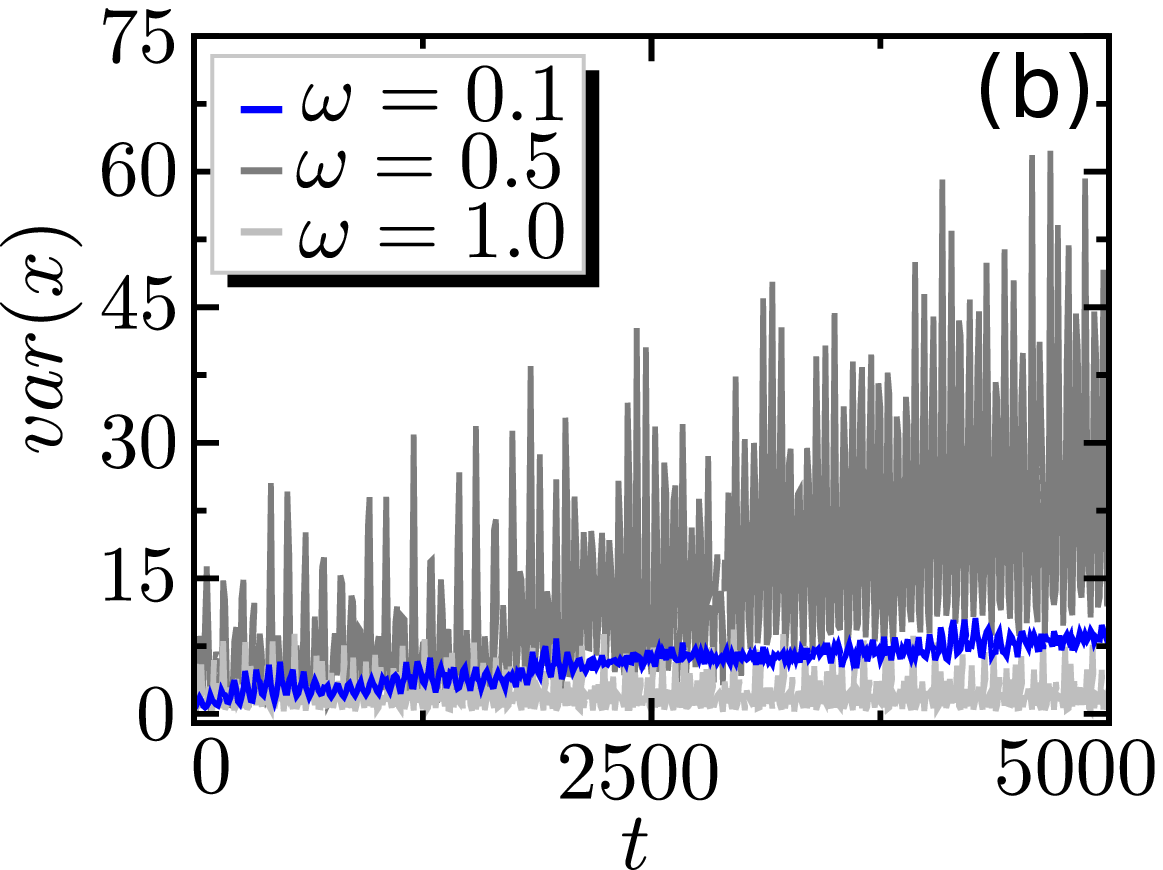}
\includegraphics[width=0.24\columnwidth]{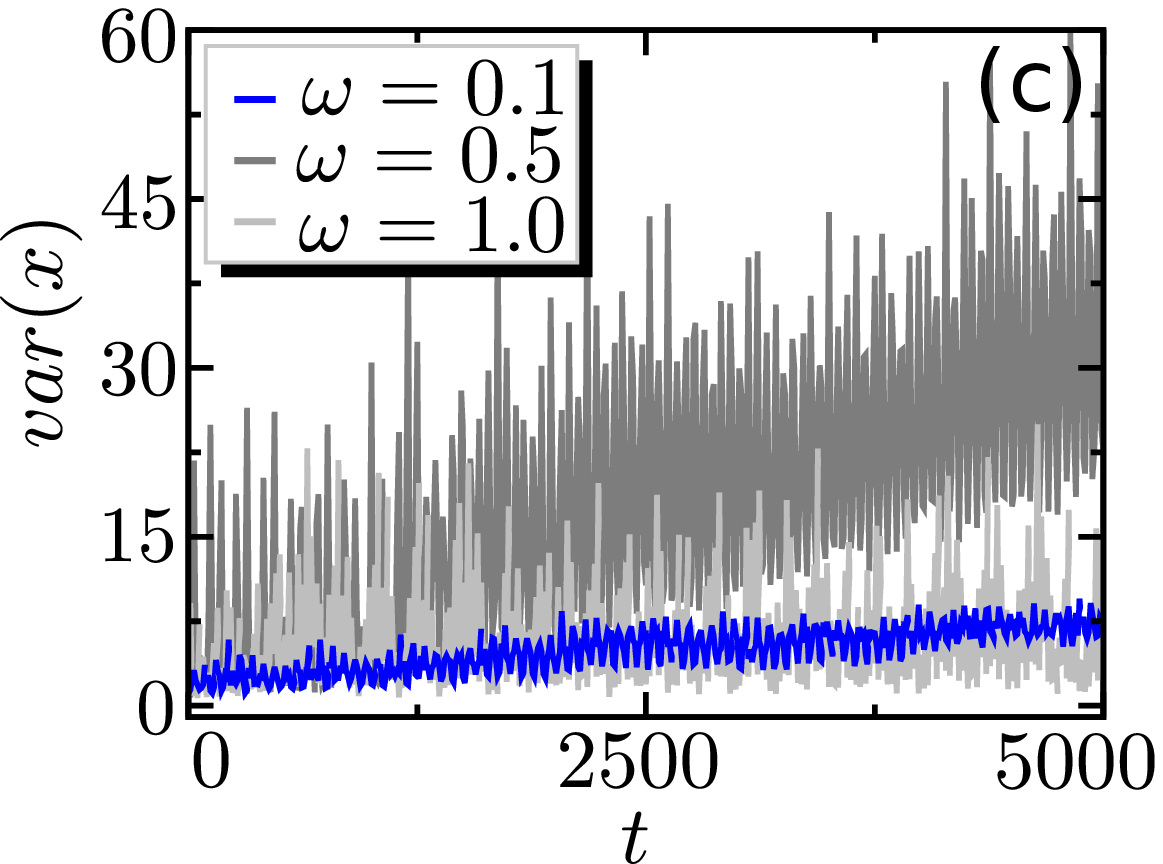} \includegraphics[width=0.24\columnwidth]{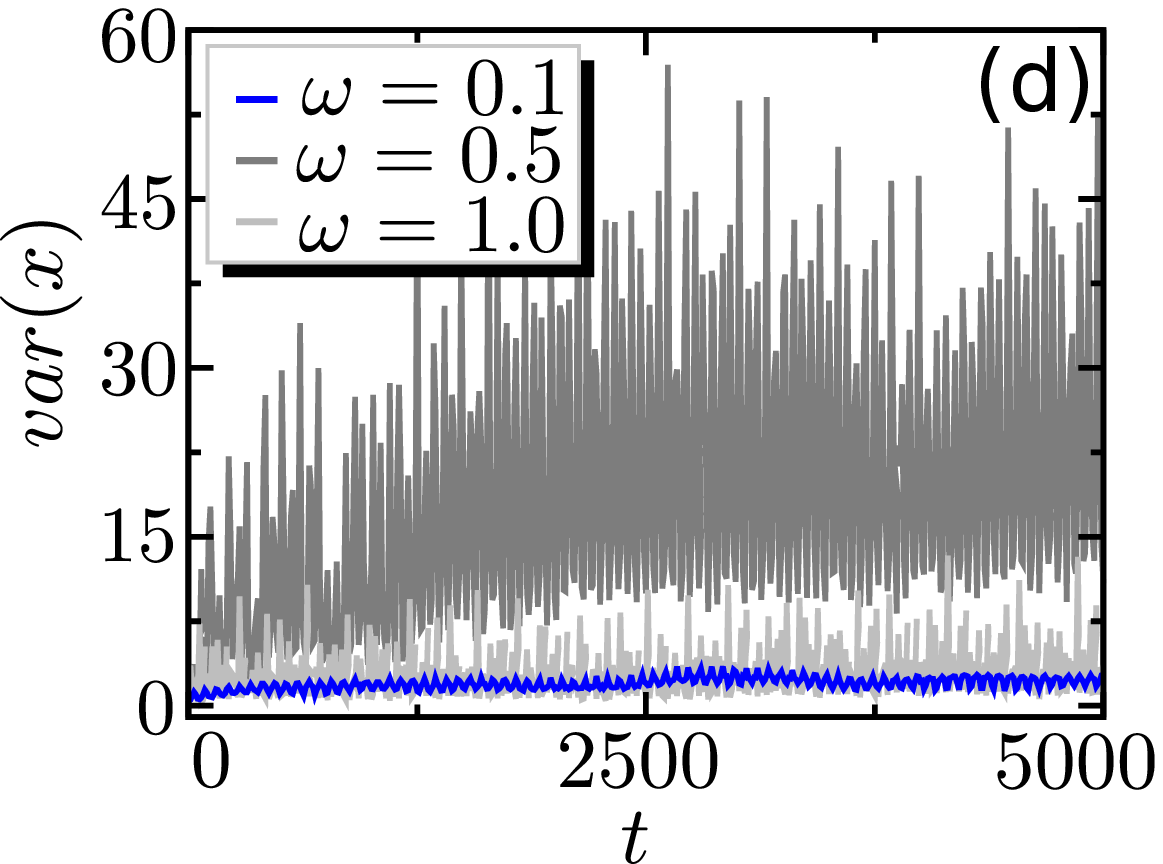}

\caption{(Color online) Stability tests via direct numerical simulations, using
the input state given by (\ref{perturbed}) with different modulation
frequencies, \emph{viz.} $\omega=0.1$, $0.5$, and $1.0$. We show
in panels (a)-(d) the variance of $x$ versus $t$ ($var(x)$)
corresponding to the cases displayed in Figs. \ref{FN7} (a)-\ref{FN7}(d),
respectively.}
\label{FN8} 
\end{figure}

%%%%%%%%%%%%%%%%%%%%%%%%%%%%%%%%%

In Figs. \ref{FN7} (a)-(d) we show the analytical profiles of the
modulated solutions by the mixed potential. Indeed, as expected we
observe the solution oscillating around the center of the trap, in a way
similar to the case of the seesaw potential, plus a breathing pattern,
as in the case of the flying-bird potential. Similarly to the case of
flying-bird potential, due to the breathing pattern of the mixed potential
the variance of $x$ will also oscillate around a constant value. This pattern was
observed in the stable solutions obtained by the numerical simulations. 

The temporal evolution of variance in $x$ of the solutions are displayed
in Figs. \ref{FN8}(a)-\ref{FN8}(d). Specifically, for the Lorentzian
solution (case A, displayed in Fig. \ref{FN8}(a)) the solution stabilizes
for $\omega\geq0.2$. For the cases B-D, we verify instability regions
similar to those observed in the flying-bird potential, i.e., the
solution loses its stability for $\omega\in[0.4,0.65]$ in case B, for
$\omega\in[0.3,0.55]$ and for $\omega=0.7$ in case C, and for $\omega\in[0.25,0.55]$
and for $\omega\in[0.75,0.8]$ in case D. In fact, one can see in the
example shown in Fig. \ref{FN8}(a) the increase in the value of the
variance in $x$ for $\omega=0.1$ while for $\omega=0.5$ and $1.0$
this parameter will oscillate around a constant value. In the same
way, in Figs. \ref{FN8}(b)-\ref{FN8}(d) the curve of $var(x)$ with
$\omega=0.5$ presents an increasing behavior illustrating the instability
region mentioned above. In all cases the temporal evolution of the
3D profiles corroborate the variance behavior.

%%%%%%%%%%%%%%%%%%%%%%%%%%%%%%%%%

\section{Summary\label{sec:Conclusion}}

In this work, we investigated the NLS equation in the presence
of quadratic and cubic nonlinearities that are modulated in time,
under the action of a background potential modulated in space and
time. We have studied two distinct solutions, constructed from Eqs.
\eqref{lorentizian} and \eqref{cosh}, and we dealt with several
possibilities, with the nonlinearities being focusing or defocusing,
and with the potential having four distinct features, as they appear
in the subsections \ref{subsec:Vanish-potential}, \ref{subsec:Seesaw-potential},
\ref{subsec:Flying-bird-potential}, and \ref{subsec:Mixed-potential}.
The study presented several analytical solutions, showing
how they can be stable or unstable, when one varies parameters
that control each one of the four specific problems considered in the work.

The main results of the current study are exemplified in the eight figures
that are depicted above. The four odd numbered figures illustrate solutions
with vanishing potential, and with potential of the seesaw, flying-bird,
or mixed type, respectively. Also, the four even figures describe the corresponding
stability, which we investigated numerically. Interestingly, we see
that the frequency of modulation is an important parameter to control
stability of the solutions for both the self-focusing and the self-defocusing cubic
nonlinearity.
Among all the interesting results, we stress that the stability is not
guaranteed for certain types of modulations, so the pattern of modulation
can work to stabilize or destabilize the solutions. 

The several investigations illustrate how to stabilize unstable solutions and how
to accomplish this possibility in a diversity of scenarios. Thus, from
the experimental point of view one can, for example, stabilize an unstable
solution, like the one shown in Fig. \ref{FN2}a, with an appropriate choice of
the pattern of modulation, and with a specific choice in the frequency of modulation,
which is the key parameter here. In this sense, the above results encourage us
to study other models, with other types of nonlinearities and solutions. We hope
to report on this in the near future.

\section*{Acknowledgments}

We thank the Brazilian agencies CAPES, CNPq and the Instituto Nacional
de Ci\^encia e Tecnologia-Informa\c c\~ao Qu\^antica (INCT-IQ) for partial
support.

\section*{References}

\bibliographystyle{vancouver}
\bibliography{Refs}

\begin{thebibliography}{10}

\bibitem{Agrawal_13}
Agrawal GP.
\newblock {Nonlinear Fiber Optics}.
\newblock Academic Press. Academic Press; 2013.
\newblock Available from:
  \url{https://books.google.com.br/books?id=xNvw-GDVn84C
  https://books.google.com.br/books?id=b5S0JqHMoxAC}.

\bibitem{Flach_PR98}
Flach S, Willis CR.
\newblock {Discrete breathers}.
\newblock Phys Rep. 1998 mar;295(5):181--264.
\newblock Available from:
  \url{http://linkinghub.elsevier.com/retrieve/pii/S0370157397000689}.

\bibitem{Andersen_PRB93}
Andersen JD, Kenkre VM.
\newblock {Self-trapping and time evolution in some spatially extended quantum
  nonlinear systems: Exact solutions}.
\newblock Phys Rev B. 1993 may;47(17):11134--11142.
\newblock Available from:
  \url{http://link.aps.org/doi/10.1103/PhysRevB.47.11134}.

\bibitem{Peyrard_PD93}
Peyrard M, Dauxois T, Hoyet H, Willis CR.
\newblock {Biomolecular dynamics of DNA: statistical mechanics and dynamical
  models}.
\newblock Phys D Nonlinear Phenom. 1993 sep;68(1):104--115.
\newblock Available from:
  \url{http://linkinghub.elsevier.com/retrieve/pii/016727899390035Y}.

\bibitem{Scott_03}
Scott A.
\newblock {Nonlinear Science: Emergence and Dynamics of Coherent Structures}.
\newblock Oxford texts in applied and engineering mathematics. Oxford
  University Press; 2003.
\newblock Available from:
  \url{https://books.google.com.br/books?id=PkxK9AVQ6SgC}.

\bibitem{Drazin_89}
Drazin PG, Johnson RS.
\newblock {Solitons: An Introduction}.
\newblock Cambridge Computer Science Texts. Cambridge University Press; 1989.
\newblock Available from:
  \url{https://books.google.com.br/books?id=HPmbIDk2u-gC}.

\bibitem{Remoissenet_13}
Remoissenet M.
\newblock {Waves Called Solitons: Concepts and Experiments}.
\newblock Springer Berlin Heidelberg; 2013.
\newblock Available from:
  \url{https://books.google.com.br/books?id=qULtCAAAQBAJ}.

\bibitem{Eilenberger_12}
Eilenberger G.
\newblock {Solitons: Mathematical Methods for Physicists}.
\newblock Springer Series in Solid-State Sciences. Springer Berlin Heidelberg;
  2012.
\newblock Available from:
  \url{https://books.google.com.br/books?id=YKvrCAAAQBAJ}.

\bibitem{Pethick_02}
Pethick CJ, Smith H.
\newblock {Bose-Einstein Condensation in Dilute Gases}.
\newblock Cambridge University Press; 2002.

\bibitem{Pitaevskii_03}
Pitaevskii LP, Stringari S.
\newblock {Bose-Einstein Condensation}.
\newblock International Series of Monographs on Physics. Clarendon Press; 2003.
\newblock Available from:
  \url{https://books.google.com.br/books?id=rIobbOxC4j4C}.

\bibitem{Malomed06-2}
Malomed BA.
\newblock {Nonlinear Schr{\"{o}}dinger Equations}.
\newblock In: Encycl. Nonlinear Sci. Taylor and Francis; 2006. p. 639--643.
\newblock Available from:
  \url{https://books.google.com.br/books?id=KC7gZmIEAiwC}.

\bibitem{Buryak_PR02}
Buryak A.
\newblock {Optical solitons due to quadratic nonlinearities: from basic physics
  to futuristic applications}.
\newblock Phys Rep. 2002 nov;370(2):63--235.
\newblock Available from:
  \url{http://linkinghub.elsevier.com/retrieve/pii/S0370157302001965}.

\bibitem{Hayata_JOSAB94}
Hayata K, Koshiba M.
\newblock {Prediction of unique solitary-wave polaritons in quadratic-cubic
  nonlinear dispersive media}.
\newblock J Opt Soc Am B. 1994 dec;11(12):2581.
\newblock Available from:
  \url{https://www.osapublishing.org/abstract.cfm?URI=josab-11-12-2581}.

\bibitem{Fujioka_CHAOS11}
Fujioka J, Cort{\'{e}}s E, P{\'{e}}rez-Pascual R, Rodr{\'{i}}guez RF, Espinosa
  A, Malomed BA.
\newblock {Chaotic solitons in the quadratic-cubic nonlinear Schr{\"{o}}dinger
  equation under nonlinearity management}.
\newblock Chaos An Interdiscip J Nonlinear Sci. 2011;21(3):033120.
\newblock Available from:
  \url{http://scitation.aip.org/content/aip/journal/chaos/21/3/10.1063/1.3629985}.

\bibitem{Dalfovo_RMP99}
Dalfovo F, Giorgini S, Pitaevskii LP, Stringari S.
\newblock {Theory of Bose-Einstein condensation in trapped gases}.
\newblock Rev Mod Phys. 1999 apr;71(3):463--512.
\newblock Available from:
  \url{http://link.aps.org/doi/10.1103/RevModPhys.71.463}.

\bibitem{Aranson_RMP02}
Aranson IS, Kramer L.
\newblock {The world of the complex Ginzburg-Landau equation}.
\newblock Rev Mod Phys. 2002 feb;74(1):99--143.
\newblock Available from:
  \url{http://link.aps.org/doi/10.1103/RevModPhys.74.99}.

\bibitem{Akhmediev_PRE96}
Akhmediev NN, Afanasjev VV, Soto-Crespo JM.
\newblock {Singularities and special soliton solutions of the cubic-quintic
  complex Ginzburg-Landau equation}.
\newblock Phys Rev E. 1996 jan;53(1):1190--1201.
\newblock Available from:
  \url{http://link.aps.org/doi/10.1103/PhysRevE.53.1190}.

\bibitem{Mihalache_PRE00}
Mihalache D, Mazilu D, Crasovan LC, Malomed BA, Lederer F.
\newblock {Three-dimensional spinning solitons in the cubic-quintic nonlinear
  medium}.
\newblock Phys Rev E. 2000 jun;61(6):7142--7145.
\newblock Available from:
  \url{http://link.aps.org/doi/10.1103/PhysRevE.61.7142}.

\bibitem{Avelar_PRE09}
Avelar AT, Bazeia D, Cardoso WB.
\newblock {Solitons with cubic and quintic nonlinearities modulated in space
  and time}.
\newblock Phys Rev E. 2009 feb;79(2):025602.
\newblock Available from:
  \url{http://link.aps.org/doi/10.1103/PhysRevE.79.025602}.

\bibitem{Cardoso_PRE11}
Cardoso WB, Avelar AT, Bazeia D.
\newblock {One-dimensional reduction of the three-dimenstional Gross-Pitaevskii
  equation with two- and three-body interactions}.
\newblock Phys Rev E. 2011 mar;83(3):036604.
\newblock Available from:
  \url{http://link.aps.org/doi/10.1103/PhysRevE.83.036604}.

\bibitem{Alfimov_PRA07}
Alfimov GL, Konotop VV, Pacciani P.
\newblock {Stationary localized modes of the quintic nonlinear
  Schr{\"{o}}dinger equation with a periodic potential}.
\newblock Phys Rev A. 2007 feb;75(2):023624.
\newblock Available from:
  \url{http://link.aps.org/doi/10.1103/PhysRevA.75.023624}.

\bibitem{Reyna_PRA14}
Reyna AS, de~Ara{\'{u}}jo CB.
\newblock {Nonlinearity management of photonic composites and observation of
  spatial-modulation instability due to quintic nonlinearity}.
\newblock Phys Rev A. 2014 jun;89(6):063803.
\newblock Available from:
  \url{http://link.aps.org/doi/10.1103/PhysRevA.89.063803}.

\bibitem{Salasnich_PRA02}
Salasnich L, Parola A, Reatto L.
\newblock {Effective wave equations for the dynamics of cigar-shaped and
  disk-shaped Bose condensates}.
\newblock Phys Rev A. 2002 apr;65(4):043614.
\newblock Available from:
  \url{http://link.aps.org/doi/10.1103/PhysRevA.65.043614}.

\bibitem{Salasnich_PRA02-2}
Salasnich L, Parola A, Reatto L.
\newblock {Condensate bright solitons under transverse confinement}.
\newblock Phys Rev A. 2002 oct;66(4):043603.
\newblock Available from:
  \url{http://link.aps.org/doi/10.1103/PhysRevA.66.043603}.

\bibitem{Mateo_PRA08}
Mateo AM, Delgado V.
\newblock {Effective mean-field equations for cigar-shaped and disk-shaped
  Bose-Einstein condensates}.
\newblock Phys Rev A. 2008 jan;77(1):013617.
\newblock Available from:
  \url{http://link.aps.org/doi/10.1103/PhysRevA.77.013617}.

\bibitem{Adhikari_NJP09}
Adhikari SK, Salasnich L.
\newblock {Effective nonlinear Schr{\"{o}}dinger equations for cigar-shaped and
  disc-shaped Fermi superfluids at unitarity}.
\newblock New J Phys. 2009 feb;11(2):023011.
\newblock Available from:
  \url{http://stacks.iop.org/1367-2630/11/i=2/a=023011?key=crossref.206d6a141e90aff3fffa0cee8a11aa07}.

\bibitem{Cardoso_PRE13}
Cardoso WB, Zeng J, Avelar AT, Bazeia D, Malomed BA.
\newblock {Bright solitons from the nonpolynomial Schr{\"{o}}dinger equation
  with inhomogeneous defocusing nonlinearities}.
\newblock Phys Rev E. 2013 aug;88(2):025201.
\newblock Available from:
  \url{http://link.aps.org/doi/10.1103/PhysRevE.88.025201}.

\bibitem{Couto_JPB15}
Couto HLC, Cardoso WB.
\newblock {Dynamics of the soliton-sound interaction in the
  quasi-one-dimensional Munoz-Mateo--Delgado equation}.
\newblock J Phys B At Mol Opt Phys. 2015 jan;48(2):025301.
\newblock Available from:
  \url{http://stacks.iop.org/0953-4075/48/i=2/a=025301?key=crossref.94b619bdc9e8159b1e066778aa3ab2d4}.

\bibitem{Biswas_CNSNS10}
Biswas A, Milovi{\'{c}} D.
\newblock {Optical solitons with log-law nonlinearity}.
\newblock Commun Nonlinear Sci Numer Simul. 2010 dec;15(12):3763--3767.
\newblock Available from:
  \url{http://linkinghub.elsevier.com/retrieve/pii/S1007570410000523}.

\bibitem{Calaca_CNSNS14}
Cala{\c{c}}a L, Avelar AT, Bazeia D, Cardoso WB.
\newblock {Modulation of localized solutions for the Schr{\"{o}}dinger equation
  with logarithm nonlinearity}.
\newblock Commun Nonlinear Sci Numer Simul. 2014 sep;19(9):2928--2934.
\newblock Available from:
  \url{http://linkinghub.elsevier.com/retrieve/pii/S1007570414000550}.

\bibitem{Soto-Crespo_PRA92}
Soto-Crespo JM, Wright EM, Akhmediev NN.
\newblock {Recurrence and azimuthal-symmetry breaking of a cylindrical Gaussian
  beam in a saturable self-focusing medium}.
\newblock Phys Rev A. 1992 mar;45(5):3168--3175.
\newblock Available from:
  \url{http://link.aps.org/doi/10.1103/PhysRevA.45.3168}.

\bibitem{Stepic_PRE04}
Stepi{\'{c}} M, Kip D, Had{\v{z}}ievski L, Maluckov A.
\newblock {One-dimensional bright discrete solitons in media with saturable
  nonlinearity}.
\newblock Phys Rev E. 2004 jun;69(6):066618.
\newblock Available from:
  \url{http://link.aps.org/doi/10.1103/PhysRevE.69.066618}.

\bibitem{Melvin_PRL06}
Melvin TRO, Champneys AR, Kevrekidis PG, Cuevas J.
\newblock {Radiationless Traveling Waves in Saturable Nonlinear
  Schr{\"{o}}dinger Lattices}.
\newblock Phys Rev Lett. 2006 sep;97(12):124101.
\newblock Available from:
  \url{http://link.aps.org/doi/10.1103/PhysRevLett.97.124101}.

\bibitem{Yang_10}
Yang J.
\newblock {Nonlinear Waves in Integrable and Nonintegrable Systems}.
\newblock Society for Industrial and Applied Mathematics; 2010.
\newblock Available from:
  \url{http://epubs.siam.org/doi/book/10.1137/1.9780898719680}.

\bibitem{Wang_PP15}
Wang L, Li M, Qi FH, Xu T.
\newblock {Modulational instability, nonautonomous breathers and rogue waves
  for a variable-coefficient derivative nonlinear Schr{\"{o}}dinger equation in
  the inhomogeneous plasmas}.
\newblock Phys Plasmas. 2015 mar;22(3):032308.
\newblock Available from:
  \url{http://scitation.aip.org/content/aip/journal/pop/22/3/10.1063/1.4915516}.

\bibitem{Theis_PRL04}
Theis M, Thalhammer G, Winkler K, Hellwig M, Ruff G, Grimm R, et~al.
\newblock {Tuning the Scattering Length with an Optically Induced Feshbach
  Resonance}.
\newblock Phys Rev Lett. 2004 sep;93(12):123001.
\newblock Available from:
  \url{http://link.aps.org/doi/10.1103/PhysRevLett.93.123001}.

\bibitem{Belmonte-Beitia_PRL08}
Belmonte-Beitia J, P{\'{e}}rez-Garc{\'{i}}a VM, Vekslerchik V, Konotop VV.
\newblock {Localized Nonlinear Waves in Systems with Time- and Space-Modulated
  Nonlinearities}.
\newblock Phys Rev Lett. 2008 apr;100(16):164102.
\newblock Available from:
  \url{http://link.aps.org/doi/10.1103/PhysRevLett.100.164102}.

\bibitem{Yan_PRE09}
Yan Z, Konotop VV.
\newblock {Exact solutions to three-dimensional generalized nonlinear
  Schr{\"{o}}dinger equations with varying potential and nonlinearities}.
\newblock Phys Rev E. 2009 sep;80(3):036607.
\newblock Available from:
  \url{http://link.aps.org/doi/10.1103/PhysRevE.80.036607}.

\bibitem{Yan_PRA09}
Yan Z, Hang C.
\newblock {Analytical three-dimensional bright solitons and soliton pairs in
  Bose-Einstein condensates with time-space modulation}.
\newblock Phys Rev A. 2009 dec;80(6):063626.
\newblock Available from:
  \url{http://link.aps.org/doi/10.1103/PhysRevA.80.063626}.

\bibitem{Avelar_PRE10}
Avelar AT, Bazeia D, Cardoso WB.
\newblock {Modulation of breathers in the three-dimensional nonlinear
  Gross-Pitaevskii equation}.
\newblock Phys Rev E. 2010 nov;82(5):057601.
\newblock Available from:
  \url{http://link.aps.org/doi/10.1103/PhysRevE.82.057601}.

\bibitem{Cardoso_NA10}
Cardoso WB, Avelar AT, Bazeia D.
\newblock {Bright and dark solitons in a periodically attractive and expulsive
  potential with nonlinearities modulated in space and time}.
\newblock Nonlinear Anal Real World Appl. 2010 oct;11(5):4269--4274.
\newblock Available from:
  \url{http://linkinghub.elsevier.com/retrieve/pii/S1468121810000805}.

\bibitem{Cardoso_PLA10}
Cardoso WB, Avelar AT, Bazeia D.
\newblock {Modulation of breathers in cigar-shaped Bose--Einstein condensates}.
\newblock Phys Lett A. 2010 jun;374(26):2640--2645.
\newblock Available from:
  \url{http://linkinghub.elsevier.com/retrieve/pii/S0375960110004895}.

\bibitem{Cardoso_PLA10-2}
Cardoso WB, Avelar AT, Bazeia D, Hussein MS.
\newblock {Solitons of two-component Bose-Einstein condensates modulated in
  space and time}.
\newblock Phys Lett A. 2010 may;374(23):2356--2360.
\newblock Available from:
  \url{http://linkinghub.elsevier.com/retrieve/pii/S0375960110004068}.

\bibitem{Serkin_PRA10}
Serkin VN, Hasegawa A, Belyaeva TL.
\newblock {Nonautonomous matter-wave solitons near the Feshbach resonance}.
\newblock Phys Rev A. 2010 feb;81(2):023610.
\newblock Available from:
  \url{http://link.aps.org/doi/10.1103/PhysRevA.81.023610}.

\bibitem{Serkin_JMO10}
Serkin VN, Hasegawa A, Belyaeva TL.
\newblock {Solitary waves in nonautonomous nonlinear and dispersive systems:
  nonautonomous solitons}.
\newblock J Mod Opt. 2010 aug;57(14-15):1456--1472.
\newblock Available from:
  \url{http://www.tandfonline.com/doi/abs/10.1080/09500341003624750}.

\bibitem{Zhang_PRA10}
Zhang JF, Tian Q, Wang YY, Dai CQ, Wu L.
\newblock {Self-similar optical pulses in competing cubic-quintic nonlinear
  media with distributed coefficients}.
\newblock Phys Rev A. 2010 feb;81(2):023832.
\newblock Available from:
  \url{http://link.aps.org/doi/10.1103/PhysRevA.81.023832}.

\bibitem{Yan_PLA10}
Yan Z.
\newblock {Nonautonomous "rogons" in the inhomogeneous nonlinear
  Schr{\"{o}}dinger equation with variable coefficients}.
\newblock Phys Lett A. 2010 jan;374(4):672--679.
\newblock Available from:
  \url{http://linkinghub.elsevier.com/retrieve/pii/S0375960109014625}.

\bibitem{He_PRE11}
He JR, Li HM.
\newblock {Analytical solitary-wave solutions of the generalized nonautonomous
  cubic-quintic nonlinear Schr{\"{o}}dinger equation with different external
  potentials}.
\newblock Phys Rev E. 2011 jun;83(6):066607.
\newblock Available from:
  \url{http://link.aps.org/doi/10.1103/PhysRevE.83.066607}.

\bibitem{He_OC12}
He Jd, Zhang Jf, Zhang My, Dai Cq.
\newblock {Analytical nonautonomous soliton solutions for the cubic--quintic
  nonlinear Schr{\"{o}}dinger equation with distributed coefficients}.
\newblock Opt Commun. 2012 mar;285(5):755--760.
\newblock Available from:
  \url{http://linkinghub.elsevier.com/retrieve/pii/S0030401811012302}.

\bibitem{Dai_AP12}
Dai CQ, Wang YY, Tian Q, Zhang JF.
\newblock {The management and containment of self-similar rogue waves in the
  inhomogeneous nonlinear Schr{\"{o}}dinger equation}.
\newblock Ann Phys (N Y). 2012 feb;327(2):512--521.
\newblock Available from:
  \url{http://linkinghub.elsevier.com/retrieve/pii/S0003491611001898}.

\bibitem{Cardoso_PRE12}
Cardoso WB, Avelar AT, Bazeia D.
\newblock {Modulation of localized solutions in a system of two coupled
  nonlinear Schr{\"{o}}dinger equations}.
\newblock Phys Rev E. 2012 aug;86(2):27601.
\newblock Available from:
  \url{http://link.aps.org/doi/10.1103/PhysRevE.86.027601}.

\bibitem{Arroyo-Meza_PRE12}
{Arroyo Meza} LE, {de Souza Dutra} A, Hott MB.
\newblock {Wide localized solitons in systems with time- and space-modulated
  nonlinearities}.
\newblock Phys Rev E. 2012 aug;86(2):026605.
\newblock Available from:
  \url{http://link.aps.org/doi/10.1103/PhysRevE.86.026605}.

\bibitem{Yomba_PLA13}
Yomba E, Zakeri GA.
\newblock {Solitons in a generalized space- and time-variable coefficients
  nonlinear Schr{\"{o}}dinger equation with higher-order terms}.
\newblock Phys Lett A. 2013 dec;377(42):2995--3004.
\newblock Available from:
  \url{http://linkinghub.elsevier.com/retrieve/pii/S0375960113008098}.

\bibitem{He_PLA13}
He JR, Yi L, Li HM.
\newblock {Localized nonlinear waves in combined time-dependent
  magnetic--optical potentials with spatiotemporally modulated nonlinearities}.
\newblock Phys Lett A. 2013 nov;377(34-36):2034--2040.
\newblock Available from:
  \url{http://linkinghub.elsevier.com/retrieve/pii/S0375960113006014}.

\bibitem{Zhong_O13}
Zhong WP, Beli{\'{c}} MR, Huang T.
\newblock {Periodic soliton solutions of the nonlinear Schr{\"{o}}dinger
  equation with variable nonlinearity and external parabolic potential}.
\newblock Opt - Int J Light Electron Opt. 2013 aug;124(16):2397--2400.
\newblock Available from:
  \url{http://linkinghub.elsevier.com/retrieve/pii/S0030402612006948}.

\bibitem{He_PLA14}
He JR, Yi L.
\newblock {Formations of n-order two-soliton bound states in Bose--Einstein
  condensates with spatiotemporally modulated nonlinearities}.
\newblock Phys Lett A. 2014 mar;378(16-17):1085--1090.
\newblock Available from:
  \url{http://linkinghub.elsevier.com/retrieve/pii/S0375960114001303}.

\bibitem{Soloman-Raju_OC15}
{Soloman Raju} T.
\newblock {Dynamics of self-similar waves in asymmetric twin-core fibers with
  Airy--Bessel modulated nonlinearity}.
\newblock Opt Commun. 2015 jul;346:74--79.
\newblock Available from:
  \url{http://linkinghub.elsevier.com/retrieve/pii/S0030401815001157}.

\bibitem{Temgoua_PRE15}
Temgoua DDE, Kofane TC.
\newblock {Nonparaxial rogue waves in optical Kerr media}.
\newblock Phys Rev E. 2015 jun;91(6):063201.
\newblock Available from:
  \url{http://link.aps.org/doi/10.1103/PhysRevE.91.063201}.

\bibitem{Yang_JMP15}
Yang Y, Yan Z, Mihalache D.
\newblock {Controlling temporal solitary waves in the generalized inhomogeneous
  coupled nonlinear Schr{\"{o}}dinger equations with varying source terms}.
\newblock J Math Phys. 2015 may;56(5):053508.
\newblock Available from:
  \url{http://scitation.aip.org/content/aip/journal/jmp/56/5/10.1063/1.4921641}.

\bibitem{Kumar-De_OC15}
{Kumar De} K, Goyal A, Raju TS, Kumar CN, Panigrahi PK.
\newblock {Riccati parameterized self-similar waves in two-dimensional
  graded-index waveguide}.
\newblock Opt Commun. 2015 apr;341:15--21.
\newblock Available from:
  \url{http://linkinghub.elsevier.com/retrieve/pii/S0030401814011444}.

\bibitem{Meza_PRE15}
Meza LEA, Dutra AdS, Hott MB, Roy P.
\newblock {Wide localized solutions of the parity-time-symmetric nonautonomous
  nonlinear Schr{\"{o}}dinger equation}.
\newblock Phys Rev E. 2015 jan;91(1):013205.
\newblock Available from:
  \url{http://link.aps.org/doi/10.1103/PhysRevE.91.013205}.

\bibitem{Rajaraman_87}
Rajaraman R.
\newblock {Solitons and Instantons: An Introd. to Solitons and Instantons in
  Quantum Field Theory}.
\newblock North-Holland; 1987.
\newblock Available from:
  \url{https://books.google.com.br/books?id=r-HCoAEACAAJ}.

\bibitem{Avelar_PLA09}
Avelar AT, Bazeia D, Cardoso WB, Losano L.
\newblock {Lump-like structures in scalar-field models in dimensions}.
\newblock Phys Lett A. 2009 dec;374(2):222--227.
\newblock Available from:
  \url{http://linkinghub.elsevier.com/retrieve/pii/S037596010901353X}.

\bibitem{Avelar_EPJC08}
Avelar AT, Bazeia D, Losano L, Menezes R.
\newblock {New lump-like structures in scalar-field models}.
\newblock Eur Phys J C. 2008 may;55(1):133--143.
\newblock Available from:
  \url{http://www.springerlink.com/index/10.1140/epjc/s10052-008-0578-6
  http://dx.doi.org/10.1140/epjc/s10052-008-0578-6}.

\bibitem{Munoz-Mateo_AP09}
{Mu{\~{n}}oz Mateo} A, Delgado V.
\newblock {Effective one-dimensional dynamics of elongated Bose-Einstein
  condensates}.
\newblock Ann Phys (N Y). 2009 mar;324(3):709--724.
\newblock Available from:
  \url{http://linkinghub.elsevier.com/retrieve/pii/S0003491608001516}.

\bibitem{Sinha_PRL07}
Sinha S, Santos L.
\newblock {Cold Dipolar Gases in Quasi-One-Dimensional Geometries}.
\newblock Phys Rev Lett. 2007 oct;99(14):140406.
\newblock Available from:
  \url{http://link.aps.org/doi/10.1103/PhysRevLett.99.140406}.

\bibitem{Kartashov_RMP11}
Kartashov YV, Malomed BA, Torner L.
\newblock {Solitons in nonlinear lattices}.
\newblock Rev Mod Phys. 2011 apr;83(1):247--305.
\newblock Available from:
  \url{http://link.aps.org/doi/10.1103/RevModPhys.83.247}.

\bibitem{Malomed_06}
Malomed BA.
\newblock {Soliton Management in Periodic Systems}.
\newblock Springer US; 2006.
\newblock Available from:
  \url{https://books.google.com.br/books?id=sr2txF_x6AgC}.

\bibitem{Kevrekidis_PRL03}
Kevrekidis PG, Theocharis G, Frantzeskakis DJ, Malomed BA.
\newblock {Feshbach Resonance Management for Bose-Einstein Condensates}.
\newblock Phys Rev Lett. 2003 jun;90(23):230401.
\newblock Available from:
  \url{http://link.aps.org/doi/10.1103/PhysRevLett.90.230401}.

\end{thebibliography}

\end{document}